\documentclass[a4paper,11pt]{article}
\pdfoutput=1 

\usepackage{jcappub} 

\usepackage[normalem]{ulem}
\usepackage[utf8]{inputenc}

\usepackage{bm}
\let\vec\bm

\newcommand{\hmpc}{\,h^{-1}\text{Mpc}}
\newcommand{\hmpci}{\,h \, \text{Mpc}^{-1}}
\newcommand{\hgpc}{\, \text{Gpc}\,h^{-1}}

\newcommand{\T}{\mathbf{\hat{\mathcal{T}}}}

\newcommand{\Dk}[1]{\frac{d^3#1}{(2\pi)^3}}

\newcommand{\vk}{\vec k}
\newcommand{\vp}{\vec p}
\newcommand{\vq}{\vec q}
\newcommand{\vx}{\vec x}
\newcommand{\vv}{\vec v}
\newcommand{\vs}{\vec s}
\newcommand{\vu}{\vec u}
\newcommand{\vhn}{\hat{\vec n}}

\newcommand{\A}{\mathcal{A}}
\newcommand{\B}{\mathcal{B}}
\newcommand{\mA}{\mathcal{A}}
\newcommand{\mB}{\mathcal{B}}

\newcommand{\tm}{m}

\newcommand{\ikk}{\underset{\vk_{12}= \vk}{\int}}
\newcommand{\ikkk}{\underset{\vk_{123}= \vk}{\int}}
\newcommand{\ip}{\int_{\vp}} 

\newcommand{\dD}{\delta_\text{D}}
\newcommand{\Ps}{\vec \Psi}

\newcommand{\revised}[1]{#1}

\title{\boldmath Clustering in Massive Neutrino Cosmologies via Eulerian Perturbation Theory}

\author[a,b]{Alejandro Aviles,}
\emailAdd{avilescervantes@gmail.com}

\author[c]{Arka Banerjee,}
\emailAdd{arka@fnal.gov}

\author[f]{Gustavo Niz,}
\emailAdd{g.niz@ugto.mx}

\author[g,h]{Zachary Slepian}
\emailAdd{zslepian@ufl.edu}

\affiliation[a]{Consejo Nacional de Ciencia y Tecnolog\'ia, Av. Insurgentes Sur 1582,
Colonia Cr\'edito Constructor, Del. Benito Ju\'arez, 03940, Ciudad de M\'exico, M\'exico}
\affiliation[b]{Departamento de F\'isica, Instituto Nacional de Investigaciones Nucleares,
Apartado Postal 18-1027, Col. Escand\'on, Ciudad de M\'exico,11801, M\'exico.}
\affiliation[c]{Fermi National Accelerator Laboratory, Cosmic Physics Center, Batavia, IL 60510, USA}
\affiliation[f]{Departamento de F\'isica, Divisi\'on de Ciencias e Ingenier\'ias, Campus Le\'on, Universidad de Guanajuato, C.P. 37150, Le\'on, M\'exico}
\affiliation[g]{Department of Astronomy, University of Florida, 211 Bryant Space Science Center, Gainesville, FL 32611, USA}
\affiliation[h]{Physics Division, Lawrence Berkeley National Laboratory, 1 Cyclotron Road, Berkeley, CA 94709, USA}

\abstract{
We introduce an Eulerian Perturbation Theory to study the clustering of tracers for cosmologies in the presence of massive neutrinos. Our approach is based on mapping recently-obtained Lagrangian Perturbation Theory results to the Eulerian framework. We add Effective Field Theory counterterms, IR-resummations and a biasing scheme to compute the one-loop redshift-space power spectrum. To assess our predictions, we compare the power spectrum multipoles against  synthetic halo catalogues from the \textsc{Quijote} simulations, finding excellent agreement on scales $k\lesssim 0.25 \hmpci$. 
\revised{One can obtain the same fitting accuracy using higher wave-numbers, but then the theory fails to give a correct estimation of the linear bias parameter.} We further discuss the implications for the tree-level bispectrum.  Finally, calculating loop corrections is computationally costly, hence we derive an accurate approximation wherein we retain only the main features of the kernels, as produced by changes to the growth rate. As a result, we show how \textsc{FFTLog} methods can be used to further accelerate the loop computations with these reduced kernels.
}

\begin{document} 
\maketitle
\flushbottom

\begin{section}{Introduction}


The standard cosmological model, which commonly assumes a massless neutrino component, has successfully explained most observations, from the exquisite final Cosmic Microwave Background (CMB) data of the Planck satellite \cite{Akrami:2018vks} to the large scale structure in the recent analysis of galaxy survey catalog release of SDSS eBOSS DR16 \cite{Alam:2020sor} or the Y3 DES results (e.g.~\cite{Pandey:2021eex}). However, the observed oscillations of atmospheric and solar neutrinos show that their flavour and mass eigenstates are not the same  \cite{Fukuda:1998mi, Ahmad:2001an, Ahmad:2002jz}, constraining the sum of neutrino masses to $\sum_i m_{\nu_i} \gtrsim 0.06 \, (0.10)$ eV for a normal (inverted) hierarchy \cite{Esteban:2018azc}.
Although particle physics experiments have reached a sub-eV sensitivity on the lightest neutrino mass (e.g. the KATRIN experiment upper bound 
$m_\nu <0.8\,\text{eV}$ (0.90 c.l.) using tritium single beta decay \cite{Aker:2019uuj,Aker:2021gma}),
the absolute mass scale, the mass hierarchy and whether neutrinos are Majorana or Dirac fermions are still open questions. 

Indeed, cosmological experiments have indirectly placed the strongest constraints on the sum of the neutrino masses, due to the non-negligible impact of these hot dark matter-like components on the overall matter evolution in the Universe. The Planck temperature-temperature power spectrum analysis constrains $\sum_i m_{\nu,i} \lesssim 0.24$ eV, which is narrowed down to $\sum_i m_{\nu,i} \lesssim 0.12$ eV when combined with Baryonic Acoustic Oscillations (BAO) data \cite{Aghanim:2018eyx}. Moreover, a ``full-shape'' analysis of the redshift-space power spectrum using the low-$z$ North Galactic Cap (NGC) data selection of the SDSS BOSS DR12 galaxy sample, without combining it with other datasets, yields $\sum_i m_{\nu,i} < 1.17$ eV (0.95 c.l.) \cite{Ivanov:2019pdj} which is still better than current particle physics experiments. Another promising avenue using higher-redshift tracers is the 1D power spectrum of the Lyman alpha forest, which places a bound of $\sum_i m_{\nu,i} < 0.71$ eV (0.95 c.l.). This bound shrinks to $\sum_i m_{\nu,i} < 0.09$ eV (0.95 c.l.) when combined with BAO data and the CMB temperature, polarization and lensing \cite{Palanque-Delabrouille:2019iyz}. It should be stressed that all the neutrino mass bounds from cosmology have strong degeneracies with other cosmological parameters.

Primordial neutrinos decouple from baryons at early epochs ($\sim \! 1 \,\text{MeV}$), when the weak interaction rate falls below the Hubble rate and is hence no longer able to maintain them in equilibrium with the primeval plasma, which at that time is composed mainly of protons, electrons and photons. Hence, neutrinos decouple while they are still relativistic, and become non-relativistic at late times, contributing to the total matter abundance in this late epoch (e.g.~\cite{Slepian:2017dld}). Despite this present-day non-relativistic nature in the bulk, neutrinos do have a large velocity dispersion, inherited from their phase-space distribution at freeze-out, which in turn prevents them from clustering below the free-streaming scale, roughly defined through a Jeans-like mechanism (e.g.~\cite{Shoji:2010hm}).

Another relevant scale in neutrino cosmology is the maximum comoving distance that neutrinos can travel over the history of the Universe. For realistic masses, this distance is about $1 \,h^{-1}\, \text{Gpc}$. Above this length-scale, neutrinos behave essentially as Cold Dark Matter (CDM); while below it, their density fluctuations are suppressed at some stage of evolution. A backreaction effect, which also damps the gravitational potentials driving the growth of structure in the Universe, leads to a suppression of CDM and baryon fluctuations as well. This implies that both the total matter and the CDM power spectra become damped below this scale. For a comprehensive review of the impact of neutrino physics in cosmology see e.g. \cite{Lesgourgues:2006nd}.

Furthermore, the matter {\it velocity} fields are also affected in the presence of massive neutrinos, mainly because the free-streaming scale induces a scale-dependent growth of overdensities, even in the linear regime. Hence the large-scale velocity field receives additional contributions compared to that in a cosmology with massless neutrinos. In particular, the Kaiser boost also becomes scale-dependent, because of the scale dependence in the logarithmic growth rate. Since the scales produced by the neutrino mass as discussed above are within the linear and quasi-linear regimes, Perturbation Theory (PT) is a valuable tool, complementary to $N$-body simulations, to study the clustering induced by massive neutrinos. In redshift-space the neutrino effects become not only more evident but also more degenerate with other PT parameters. In particular, we see this when looking at the broadband shape of the power spectrum using a multipole expansion or wedges. In this context, higher-order statistics may help to break these degeneracies (see e.g. \cite{Hahn:2019zob, Hahn:2020lou, Kamalinejad:2020izi}).

As the forthcoming generation of higher-precision galaxy surveys, such as DESI \cite{Aghamousa:2016zmz},\footnote{\href{https://www.desi.lbl.gov/}{desi.lbl.gov/}} Euclid \cite{laureijs2011euclid}\footnote{\href{https://sci.esa.int/web/euclid}{sci.esa.int/web/euclid}} or the Vera C. Rubin Observatory \cite{Ivezic:2008fe}\footnote{\href{https://www.lsst.org/}{lsst.org/}} are expected to reach enough precision to measure the absolute mass scale of the neutrinos, and potentially also its hierarchy, exhaustive analytical and semi-analytical methods should be accordingly developed for cosmological parameter inference. This was recently done using the BOSS DR12 data in \cite{Ivanov:2019pdj,DAmico:2019fhj,Colas:2019ret} with the use of EPT and the Effective Field Theory of Large-Scale Structure (EFT) 

In the current work, we construct an Eulerian Perturbation Theory (EPT) for the clustering of tracers in redshift space in the presence of massive neutrinos, which is self-consistent and well-behaved in both the UV and the IR. Our modelling is based on the Lagrangian Perturbation Theory (LPT) constructed recently by two of us in \cite{Aviles:2020cax}.  
More precisely, the velocity and density kernels are obtained by mapping the LPT kernels of \cite{Aviles:2020cax} to the Eulerian framework. Furthermore, the mapping method heavily relies on a previous study \cite{Aviles:2020wme}, where the redshift-space power spectrum with generalized kernels, beyond Einstein-de Sitter (EdS), was obtained for representative modified gravity models. The latter work is, in turn, a generalization of the velocity moments expansion approach, in the form presented in \cite{Vlah:2018ygt}. 
In addition, we include the non-linear effects of large-scales in terms of bulk flows via the IR-resummations techniques of \cite{Senatore:2014via,Baldauf:2015xfa,Ivanov:2018gjr}, which are not captured by the EPT and serve to damp the BAO features in the power spectrum. We also add EFT counterterms to model the small-scales physics, which are out of reach from the PT, together with the damping of the density and velocity fields along the line-of-sight due to Redshift Space Distortions (RSD). Finally, we use the biasing scheme of \cite{McDonald:2009dh}, slightly modified in \cite{Aviles:2020wme}, to account for additional scales. In summary and to our knowledge, this is the first work that contains all the above ingredients in redshift-space for massive neutrino cosmologies. 

One important disadvantage of the method we have described is that computing the loop corrections is computationally very costly; which makes it highly unlikely that our theory, as it stands, can be used for efficient cosmological parameter estimation in the near future. The reason for this slowness is that we have to solve several ordinary differential equations at each volume element in the loop integrals. However, we show that the main differences of the perturbative kernels from those in a $\Lambda$CDM cosmology come from the scale-dependent rate. Hence, we simplify them in a manner where the growth rate effects are maintained, and obtain simple analytical expressions that allow the use of \textsc{FFTLog} methods \cite{McEwen:2016fjn,Schmittfull:2016jsw,Simonovic:2017mhp}, dramatically reducing the computational time.  

At the same time as analytic methods have been improved, significant progress has been made in capturing the effects of massive neutrinos on structure formation down to fully nonlinear scales using $N$-body simulations. Various numerical techniques have been developed over the years to account for neutrino effects in these simulations \cite{Brandbyge_2009, Viel_2010,Ali_Ha_moud_2012,Banerjee:2016zaa,Inman_2017,Banerjee_2018,hybrid,Dakin_2019,Bayer_2020}. These techniques range from treating neutrinos as an Eulerian fluid on a grid coupled to the full nonlinear gravitational potential, adding neutrinos as a separate set of $N$-body particles, and hybrid techniques. 

Despite the wide variety of techniques used in these simulations, the results on the quasi-linear scales of interest in this paper are in agreement with each other at the percent level. These simulation results serve two purposes for various analytical and semi-analytical methods---they can be used to calibrate the analytic methods, e.g. to help fix values of certain free parameters in the model, and to validate the results from the semi-analytical methods over a wide range of the parameter space of interest. Therefore, we compare the multipoles of the power spectrum obtained from our theory against the dark matter halos obtained from the \textsc{Quijote}\footnote{\href{https://github.com/franciscovillaescusa/Quijote-simulations}{github.com/franciscovillaescusa/Quijote-simulations}} suite of simulations \cite{Villaescusa-Navarro:2019bje} at redshifts $z=0.5$ and $1.0$, and for neutrinos with total mass $M_\nu = \sum_i m_{\nu,i}=$ 0, 0.1, 0.2 and 0.4 eV, where the masses are equally distributed among the three mass eigenstates. We find excellent agreement up to scales $k \simeq 0.25 \hmpci$ for the monopole and quadrupole, as is expected generally for EFT theory. The agreement can be extended to smaller scales at the cost of losing accuracy at small wave-numbers, 
and probably such small scales are not in the range where the broadband shape of the power spectrum provides cosmological information

We would like to end this introductory section by referring the reader to other relevant works which include the effects of neutrinos using PT, certainly a topic much less explored than the $\Lambda$CDM case. We start by mentioning the early work of \cite{Hu:1997vi} which describes the linear theory only, but provides several meaningful physical insights and approximate analytical formulae for the linear growth and transfer functions. To our knowledge, the first study of nonlinear PT was done in \cite{Saito:2008bp} with the use of Einstein-de Sitter (EdS) kernels, with the subsequently refined work of  \cite{Wong:2008ws,Saito:2009ah,Shoji:2009gg,Lesgourgues:2009am,Upadhye:2013ndm,Dupuy:2013jaa,Blas:2014hya,Fuhrer:2014zka,Levi:2016tlf,Wright:2017dkw,Senatore:2017hyk,Garny:2020ilv}, which included a more complete description of non-linearities. We will highlight some of these references in what follows.

The rest of this paper is organized as follows. In  \S\ref{sect:NeutrinoDensity_LPT} we describe our treatment of linear and non-linear neutrino density fluctuations.  \S\ref{sect:LPTtoSPT} is devoted to obtaining the EPT kernels by mapping the LPT ones obtained in \cite{Aviles:2020cax}. In \S\ref{sect:1dimSpectra} we present the theory for the 1-dimensional spectra of density-density and velocity-velocity fluctuations, as well as the density-velocity cross spectra; we also present the biasing scheme. In \S\ref{sect:RSDpk}, we outline the redshift-space description of the power spectrum and add EFT counterterms and IR resummations. We confront our predictions with $N$-body simulations in  \S\ref{sect:results}. In \S\ref{sect:FFTlog} we show how the FFTLog method can be used to accelerate the loop integrals. We summarize and conclude in \S\ref{sect:conclusions}. In the appendix \ref{app:2EPTKernels} we perform a direct calculation of the second-order Eulerian kernels, complementary to the mapping performed in \S\ref{sect:LPTtoSPT}.

\end{section}

\begin{section}{Neutrino density}\label{sect:NeutrinoDensity_LPT}

Our starting point is the Poisson equation in the presence of massive neutrinos and at late times when relativistic components can be neglected, which can be written as
\begin{equation}
  -\frac{k^2}{a^2} \Phi(\vk,t) = A(k,t) \delta_{cb}(\vk,t), 
\end{equation}
where $\Phi$ is the gravitational potential and we have defined the following quantities:
\begin{equation}\label{Akt}
A(k,t) \equiv A_0 \big[ f_{cb} + f_\nu \alpha(k,t) \big], \qquad A_0\equiv \frac{3}{2} \Omega_m H^2, \qquad  \alpha(k,t) \equiv \frac{\delta_{\nu}(\vk,t)}{\delta_{cb}(\vk,t)}.
\end{equation}
Therefore, $A(k,t)$ acts as a scale-dependent gravitational strength and $A_0$ is its limit as $k\rightarrow 0$.
$\delta_{cb}$ and $\delta_\nu$ are the density fluctuations of the combined CDM + baryon fluid and the massive neutrinos, respectively, while $f_{cb} = \Omega_{cb}/\Omega_{m}$ and $f_{\nu} = \Omega_{\nu}/\Omega_{m}=1 - f_{cb}$ are the relative abundances with respect to the total matter fluid $m=cb+\nu$. We are working on sufficiently large scales that baryonic effects are negligible, and so we treat the baryons as part of the cold dark matter component; see, e.g., \cite{Chisari:2019tus}. We note that \revised{at least in linear theory} $\delta_{cb}$ and $\delta_{\nu}$ have the same phase, inherited from the curvature potential at the end of inflation, hence $\alpha(k,t)$ reduces to a function of $k$ only.


The key approximation in our analysis is to take
\begin{equation} \label{alphaApprox}
\alpha(k,t) \approx  \frac{T_{\nu}(k,t)}{T_{cb}(k,t)},  \end{equation}
where $T_\nu$ and $T_{cb}$ are the transfer functions connecting overdensities imprinted at the end of inflation with linear overdensities at time $t$. 
This approximation has been tested using simulations in \cite{Aviles:2020cax}, finding a very good agreement up to well inside the non-linear regime $k\sim 1 \hmpci$, \revised{for neutrino masses up to $M_\nu=0.4\,\text{eV}$. In more detail, the total matter power spectrum can be written as the sum 
\begin{equation*}
    P_{mm}(k)=f_{cb}^2 P_{cb}(k) + 2 f_\nu f_{cb} P_{cb,\nu}(k)+ f_\nu^2 P_{\nu}(k),
\end{equation*}
and using the approximation of eq.~\eqref{alphaApprox}, we can substitute  
\begin{align*}
   P_{cb,\nu}(k) =\alpha(k) P_{cb}(k) \quad &\longrightarrow \quad \frac{T_\nu(k)}{T_{cb}(k)} P_{cb}(k), \\
   P_{\nu}(k) =\alpha^2(k) P_{cb}(k) \quad &\longrightarrow \quad \left(\frac{T_\nu(k)}{T_{cb}(k)}\right)^2 P_{cb}(k), 
\end{align*}
to obtain 
\begin{equation*}
    P_{mm}(k)=\left[f_{cb}^2 +  2 f_\nu f_{cb} \frac{T_\nu(k)}{T_{cb}(k)} +  f_\nu^2 \left(\frac{T_\nu(k)}{T_{cb}(k)}\right)^2 \right] P_{cb}(k).
\end{equation*}
The analysis of \cite{Aviles:2020cax} extracts the $P_{mm}^\text{sims}$ and $P_{cb}^\text{sims}$ power spectra directly from the \textsc{Quijote} simulations \cite{Villaescusa-Navarro:2019bje}, and constructs an approximation of $P_{mm}^\text{approx}$ by first, substituting in the above equation $P_{cb} \rightarrow P_{cb}^\text{sims}$ and second, using the linear transfer functions computed with the code \texttt{CAMB}. A comparison to the simulated matter power spectrum yields $|P_{mm}^\text{approx}/P_{mm}^\text{sims}-1| <0.001$.} This result is consistent with previous work \cite{Lesgourgues:2009am,Upadhye:2013ndm}, which find a $0.1\%$ error in the real-space power spectrum when considering the approximation $\delta_\nu = (\delta^{(1)}_\nu/\delta^{(1)}_{cb})\delta_{cb}$ for non-linear neutrino overdensities; in agreement with our approximation [eq.~\eqref{alphaApprox}] for adiabatic perturbations, which we assume throughout this work. 

These equations show the main aspects of neutrino clustering. Neutrinos become non-relativistic when their temperature drops below their mass; this happens at redshift $1+z_{nr,i}\approx 1894 \, m_{\nu,i}/\text{eV}$ for each mass eigenstate $m_{\nu,i}$.
Afterwards, they behave as hot dark matter with abundance 
\begin{equation}
\Omega_{\nu0} = \frac{M_\nu}{93.14 \,  h^2 \, \text{eV}}, \qquad M_\nu = \sum_{i=1}^3 m_{\nu,i},   
\end{equation}
that do not cluster below their free-streaming scales, which is given for each neutrino species by \cite{Shoji:2010hm} 
\begin{equation}\label{kFS}
k_\text{FS,i}(z) \approx 0.0908  \frac{H(z)}{(1+z)^2} \left(\frac{m_{\nu,i}}{\text{0.1 eV}}\right) \hmpci.
\end{equation}
For equal mass neutrinos, $m_{\nu,i}=M_\nu/3$, the different free-streaming scales reduce to a single value, that we simply call $k_\text{FS}$. $N$-body simulations with massive neutrinos usually make this assumption, as we also consider throughout this work, even though our results do not depend on this choice. 
At very large scales, $k\ll k_\text{FS}$, neutrinos behave as CDM ($T_\nu\sim T_{cb}(k)$) and the Poisson equation is that of $\Lambda$CDM.  
In the opposite limit ($k\gg k_\text{FS}$), one gets that $T_\nu\ll T_{cb}(k)$ and the source of the gravitational potential becomes proportional to $f_{cb}$, which ends up in the damping of fluctuations at scales smaller than the free-streaming scale.

To find the evolution of $cb$ fields in PT, we must supplement the Poisson equation with the continuity and Euler equations and solve them order by order to obtain the kernels $F_n$ and $G_n$, leading to the $n$-ordered density fluctuation and velocity fields
\begin{align}
 \delta_{cb}^{(n)}(\vk,t) &= \underset{\vk_{1\cdots n}= \vk}{\int} F_n(\vk_1,\cdots,\vk_n;t)\delta_{cb}^{(1)}(\vk_1,t)\cdots \delta_{cb}^{(1)}(\vk_n,t), \label{dcbExp}\\
 \theta_{cb}^{(n)}(\vk,t) &= \underset{\vk_{1\cdots n}= \vk}{\int} G_n(\vk_1,\cdots,\vk_n;t)\delta_{cb}^{(1)}(\vk_1,t)\cdots \delta_{cb}^{(1)}(\vk_n,t), \label{tcbExp}
\end{align}
where we have adopted the shorthand notations
\begin{equation}
 \underset{\vk_{1\cdots n}= \vk}{\int} 
 = \int \Dk{\vk_1}\cdots \Dk{\vk_n}(2\pi)^3 \dD(\vk_{1\cdots n}-\vk)\quad \mathrm{ and }\quad  \vk_{1\cdots n} = \vk_1 + \cdots + \vk_n.
\end{equation}
The $\theta_{cb}$ field is the divergence of the peculiar velocity $\vv$; which more precisely we define as
\begin{equation}\label{thetacb}
 \theta_{cb}(\vk,t) = -\frac{i \vk \cdot \vv}{a H f_0},
\end{equation}
where $f$ is the logarithmic derivative of the linear growth rate $D_+$ with respect to scale factor $a$, 
\begin{equation}
 f(k,t)= \frac{d \ln D_+(k,t)}{d \ln a(t)},
\end{equation}
and 
\begin{equation}
 f_0(t) \equiv f(k\rightarrow 0,t) = f^{M_\nu=0}(t)   
\end{equation}
is the large scale limit of the growth rate, which is not sensitive to the neutrino mass because on these scales neutrinos behave as cold dark matter. One should notice that both, the linear growth function and the growth rate, become scale-dependent due to the additional scale introduced by the free-streaming. Moreover, the linear growth function $D_+$ is the solution to the equation
\begin{equation}
\ddot{D}(k,t) + 2 H \dot{D} - A(k,t) D = 0,   
\end{equation}
with the appropriate initial conditions to isolate the growing solution, which are chosen well inside the matter dominated epoch by the approximated analytical formula of \cite{Hu:1997vi}. 

In figure \ref{fig:Dp_fk}, we show plots for the scale-dependent linear growth function and growth rate of the $cb$ fluid, at redshifts $z=0$ and $z=1$, for degenerate neutrinos with total mass $M_\nu =$ 0.1, 0.2 and 0.4 eV, which for $\Omega_m=0.3175$ and $h=0.6711$ correspond to  $f_\nu=0.0075,\, 0.0150,\, 0.0300$, respectively. The linear transfer functions $T_\nu(k,z)$ and $T_{cb}(k,z)$ are computed with the code \texttt{CAMB}\footnote{\href{https://camb.info/}{https://camb.info/}.} \cite{Lewis:1999bs}. Alternatively, these function may be obtained to a very high accuracy with the analytical formulae of \cite{Hu:1997vi}.

 \begin{figure}
 	\begin{center}
 	\includegraphics[width=6.0 in]{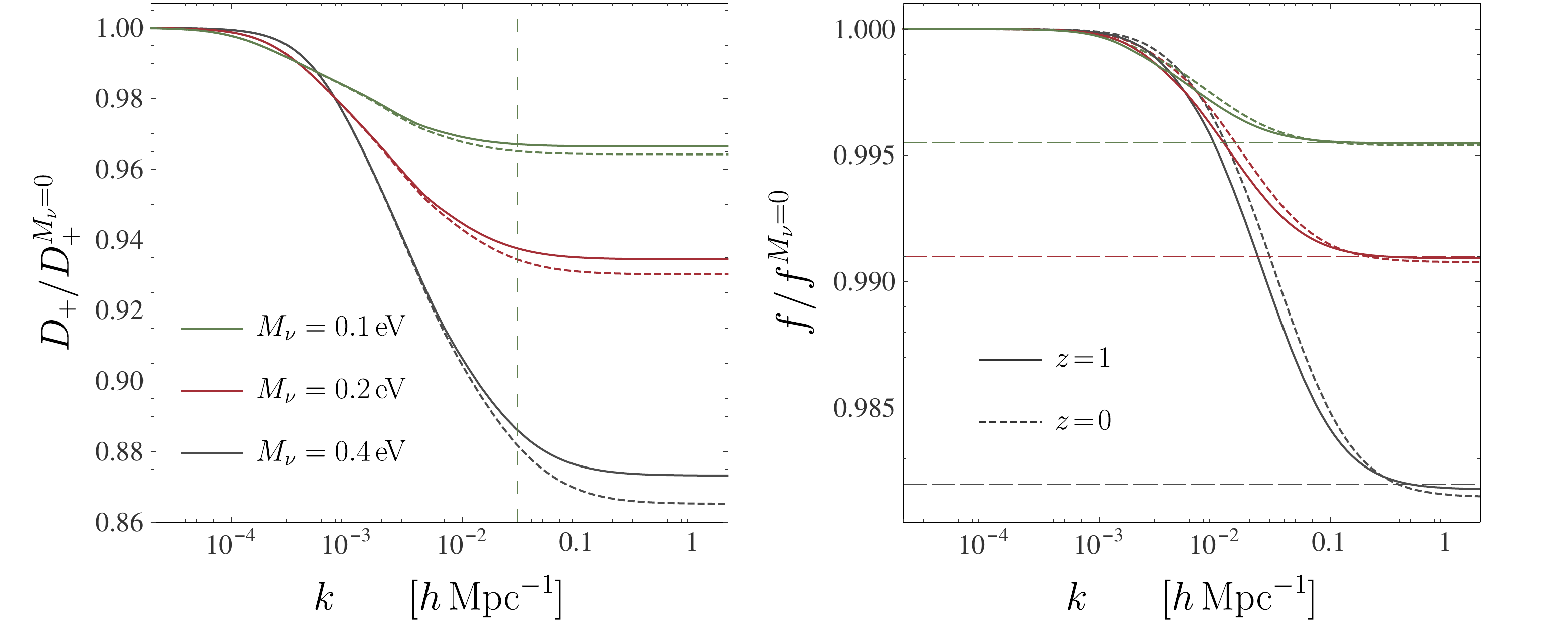}
 	\caption{Linear growth function $D_+(k,z)$ (left panel) and growth rate $f(k,z)$ (right panel) for neutrino masses $M_\nu=0.1, 0.2, 0.4$ eV, divided by the massless neutrino cases. We consider the redshifts $z=1$ (solid lines) and $z=0$ (dashed lines). In the left panel, the vertical dashed lines show the free-streaming scales at $z=0$ computed from eq.~\eqref{kFS}. In the right panel, the dashed horizontal lines show the small scales limit $f(k\gg k_\text{FS}) \approx (1-3/5 f_\nu) f^{M_\nu=0}$.    
 	\label{fig:Dp_fk}}
 	\end{center}
 \end{figure}

Using the linear growth rate $D_+$, the linear density fields evolve as 
\begin{equation}
 \delta^{(1)}_{cb}(\vk,t) = \frac{D_+(k,t)}{D_+(k,t_0)} \delta^{(1)}_{cb}(\vk,t_0),   
\end{equation}
with $t_0$ the present time. However, note that this decomposition is somewhat arbitrary since a full factorization into scale-dependent and time-dependent pieces is not possible. On the other hand, the velocity field and the density fluctuations are related by
\begin{equation}\label{lineartheta}
 \theta_{cb}^{(1)}(\vk,t) = \frac{f(k,t)}{f_0(t)} \delta_{cb}^{(1)}(\vk,t),
\end{equation}
thus coinciding at large scales but resulting in suppressed velocities, by a factor of $f(k)/f_0$,  below the free-streaming scale. From above's equation and eq.~\eqref{tcbExp} we obtain the first order kernels 
\begin{equation}
F_1(\vk,t)=1, \qquad \text{and} \qquad G_1(\vk,t)=\frac{f(k,t)}{f_0(t)}.
\end{equation}
Since the relation between the linear velocity and density fields is multiplicative in Fourier space, they become non-locally related in configuration space. However, the $f(k)$ growth rate is close to a step function, sharply interpolating in $\log(k)$, at a scale $\sim k_\text{FS}$, between $f_0$ and $f_{\infty}=f(k \rightarrow \infty)$.  
Hence, the non-local effect arising from convoluting the inverse Fourier transform of $f(k)$ with the configuration space $cb$ overdensity is small. 
\revised{In practice, the density fields are always smoothed over certain scale; and therefore, if the width of support of $\tilde{f}$ (or the physical region where it is reasonable different from zero) turns out to be smaller than the smoothing scale of the density fields, the relation between the $\theta^{(1)}_{cb}$ and $\delta^{(1)}_{cb}$ becomes effectively local.} 

Finding the higher order kernels is quite cumbersome using fluid equations and eqs.~\eqref{dcbExp} and \eqref{tcbExp}. However, in this work we take a different approach, which maps the LPT kernels found in \cite{Aviles:2020cax} to the required EPT kernels. Before presenting this mapping (in section \ref{sect:LPTtoSPT}) let us review the basic ingredients of the LPT formalism. Moreover, in appendix \ref{app:2EPTKernels}, the reader may find the second order kernels $F_2$ and $G_2$ directly.

\begin{subsection}{Lagrangian evolution}

In a Lagrangian description the observer follows  trajectories of the $cb$ fluid elements that have initial coordinates $\vq$ and final, Eulerian coordinates 
\begin{equation} \label{qTox}
 \vx(\vq,t) = \vq + \Ps(\vq,t),
\end{equation}
with $\Ps$ the Lagrangian displacement of the $cb$ field, which obeys the evolution equation
\begin{equation}\label{LDEvEq}
 \ddot{\Ps}(\vq,t) + 2 H \dot{\Ps}(\vq,t) = -\nabla \Phi(\vx,t)\Big|_{\vx=\vq+\Ps}.
\end{equation}
The non-linear neutrino fluctuations in the Lagrangian space become \cite{Aviles:2020cax}
\begin{align} \label{deltanuProxy}
\tilde{\delta}_\nu(\vk) 
&= \alpha(k)  \tilde{\delta}_{cb}(\vk) + \frac{1}{A_0 f_\nu} \ikk \mathcal{K}^\text{FL$\Psi$}_{li}(\vk_1,\vk_2) \Psi_l(\vk_1) \Psi_i(\vk_2) \nonumber\\
&\quad  + \frac{1}{A_0 f_\nu} \ikkk \mathcal{K}^\text{FL$\Psi$}_{lij}(\vk_1,\vk_2,\vk_3) \Psi_l(\vk_1) \Psi_i(\vk_2)\Psi_j(\vk_3) + \cdots,
\end{align}
where we have omit to write the time dependence. 
\revised{In general, we do not write a tilde over quantities transformed to Fourier space, since it usually does not lead to confusions. However, in the equation above we do write a tilde over the $\nu$ and $cb$ overdensities to point out that they are Fourier transforms of Eulerian quantities but with respect to the Lagrangian coordinates, that is, $\tilde{\delta}_{cb,\nu}(\vk) = \int d^3 q \, e^{-i\vk \cdot \vq} \delta_{cb,\nu}(\vx)= \int d^3 q \, e^{-i\vk \cdot \vq} \delta_{cb,\nu}(\vq + \Ps)$. In contrast, the function $\alpha(k)$ (without a tilde) is indeed the $x$-Fourier transform of the neutrino to $cb$ fluid overdensities ratio in the Eulerian framework, and not $\tilde{\alpha}(k)= \tilde{\delta}_\nu(\vk)/\tilde{\delta}_{cb}(\vk)$. As a result, eq.~\eqref{deltanuProxy} shows the non-linear nature of the relation between neutrino and $cb$ overdensities in Lagrangian space, regardless if one takes $\alpha$ to be equal to the linear transfer functions ratio; for further details and a detailed derivation we refer the reader to \cite{Aviles:2020cax}. Actually, this non-linear relation is encoded in the  kernels $\mathcal{K}^{\text{FL} \Psi}$, namely}
\begin{align}
 \mathcal{K}^\text{FL$\Psi$}_{li}(\vk_1,\vk_2) &=  \big[A(k_1)-A(k_{12})\big]  k_1^l k_1^i, \\
 \mathcal{K}^\text{FL$\Psi$}_{lij}(\vk_1,\vk_2,\vk_3) &= -i \big[A(k_{123})-A(k_1) \big] k_1^l k_1^i k_1^j \nonumber\\
 &\quad    +  i \big[ A(k_{123}) - A(k_{13}) \big] k_{13}^i \left[ k_1^j k_1^l +\frac{1}{2} k_1^l k_3^j + \frac{1}{2}k_1^j k_3^l \right],            
\end{align}
\revised{which becomes exactly zero when the function $A(k)$ (or $\alpha(k)$) is scale-independent. This is expected since it would mean no new scale was introduced into the theory, as in the limit of zero neutrino masses where the relation of eq.~(\ref{deltanuProxy}) becomes linear. 
These kernels were introduced and named ``frame-lagging'' in \cite{Aviles:2017aor}, in a slightly different mathematical form, to correct the ``lag'' between spatial derivatives taken in the Eulerian and Lagrangian frames.}

\revised{Now, the} LPT \revised{formalism} expands the Lagrangian displacements $\Ps=\Ps^{(1)} + \Ps^{(2)} + \Ps^{(3)} + \cdots$ and solves eq.~\eqref{LDEvEq} iteratively, to find 
\begin{equation}\label{LDkernels}
 \Psi_i^{(n)}(\vk,t) = \frac{i}{n!} \underset{\vk_{1\cdots n}= \vk}{\int} L^{(n)}_i(\vk_1,\cdots,\vk_n;t)\delta_{cb}^{(1)}(\vk_1,t)\cdots \delta_{cb}^{(1)}(\vk_n,t). 
\end{equation}
The LPT kernels $\vec L^{(n)}$ up to third order for massive neutrino cosmologies are found in eqs.~(3.4), (3.8) and (3.18) of \cite{Aviles:2020cax}. It is common to assume that the linear fields are purely longitudinal in these computations, so that the transverse pieces, of the form $\Ps^\text{trans}(\vk)=\vk \times \vec T$, appear at terms with orders $n>2$ \cite{Matsubara:2015ipa}. However,  in one-loop 2-point statistics, these transverse parts are contracted with external wave-vectors, $\vk \cdot \Ps^\text{trans}$, so that they do not contribute. Therefore, without loss of generality we can use the divergence of the Lagrangian fields rather than the Lagrangian displacements. Doing this will ease the computations shown in the next section. To order $n$, we have
\begin{equation}\label{LDDkernels}
 -i\vk \cdot \Ps^{(n)}(\vk,t) = \frac{1}{n!} \underset{\vk_{1\cdots n}= \vk}{\int} C^{(n)} \Gamma_n(\vk_1,\cdots,\vk_n;t)\delta_{cb}^{(1)}(\vk_1,t)\cdots \delta_{cb}^{(1)}(\vk_n,t),
\end{equation}
where $C^{(n)} \Gamma_n$ is the $n$th order kernel for $-i\vk \cdot \Ps$, and the $C^{(n)}$ are constant coefficients that we fixed, as in \cite{Valogiannis:2019nfz}, to the values $C^{(1)}=C^{(3)}=1$, and $C^{(2)}= 3/7$. Moreover, eqs.~\eqref{LDkernels} and \eqref{LDDkernels} imply
\begin{equation}
 C^{(n)} \Gamma_n(\vk_1,\cdots,\vk_n;t) = k^i_{1\cdots n} L_i^{(n)}(\vk_1,\cdots,\vk_n;t).
\end{equation}
Equivalently, for the time derivative of $\Ps$ we find at order $n$ that
\begin{equation}\label{LDDdkernels}
 -i\vk \cdot \dot{\Ps}^{(n)}(\vk,t) = \frac{1}{n!} n H f_0 \underset{\vk_{1\cdots n}= \vk}{\int} C^{(n)} \Gamma^f_n(\vk_1,\cdots,\vk_n;t)\delta_{cb}^{(1)}(\vk_1,t)\cdots \delta_{cb}^{(1)}(\vk_n,t),
\end{equation}
with
\begin{equation}
 \Gamma^f_n(\vk_1,\cdots,\vk_n;t) =  \Gamma_n(\vk_1,\cdots,\vk_n;t) \frac{f(k_1)+\cdots+f(k_n)}{n f_0} + \frac{1}{n f_0 H} \dot{\Gamma}_n(\vk_1,\cdots,\vk_n;t) .
\end{equation}
In the particular case of EdS kernels, one gets that $\Gamma^f_n=\Gamma_n$, recovering the known relation $\dot{\Ps}^{(n)} = n H f \Ps^{(n)}$ \cite{Matsubara:2007wj}. 
\end{subsection}

\end{section}

\begin{section}{From Lagrangian to Eulerian kernels}\label{sect:LPTtoSPT}

In this section we map the LPT kernels of \cite{Aviles:2020cax} to the EPT counterparts, to obtain the $F_{2,3}$ and $G_{2,3}$ kernels that we use in this analysis. We start with the Eulerian-space overdensity written in terms of Lagrangian displacements:
\begin{equation}\label{dcb_LF}
\delta_{cb}(\vk) = \int d^3q\; e^{-i\vk\cdot \vq}\Big[ e^{-i \vk \cdot \Ps(\vq)} - 1 \Big]. 
\end{equation}
To bring the above equation into a complete EPT description, in the form given in eq.~\eqref{dcbExp}, we first expand the Lagrangian displacement out of the exponential, as 
\begin{align}
\delta_{cb}(\vk) &= \sum_{m=1}^\infty \frac{(-i)^m}{m!} \int d^3q\; e^{-i\vk\cdot \vq} \big[\vk \cdot \Ps(\vq)\big]^m \\
&=  \sum_{m=1}^\infty \frac{(-i)^m}{m!} k_{i_1}\cdots  k_{i_m} \underset{\vp_{1 \cdots m}= \vk}{\int} \Psi_{i_1}(\vp_1)\cdots \Psi_{i_m}(\vp_m),
\end{align}
where in the second equality we have Fourier-transformed each Lagrangian displacement, $\Psi_i(\vq)= \int_{\vp} e^{i\vp\cdot \vk} \Psi_i(\vp)$, and performed the $\vq$ integral analytically, which yields to a Dirac delta function ensuring momentum conservation $\vp_{1}+\cdots +\vp_{m}=\vk$. The next step is to expand the Lagrangian displacements perturbatively as $\Psi_i(\vp) = \Psi_i^{(1)}(\vp)+\cdots = [\vp \cdot \Ps^{(1)}(\vp)] p_i/p^2 + \cdots$; then use eq.~\eqref{LDDkernels}, and finally relate the $\Gamma_n$ kernels with the $F_n$ kernels using eq.~\eqref{dcbExp}. This procedure yields to the following results for $F_2$ and $F_3$
\begin{align}
 F_2(\vk_1,\vk_2) &= \frac{3}{14} \Gamma_2(\vk_1,\vk_2) + \frac{1}{2}\frac{(\vk_{12}\cdot\vk_1)(\vk_{12}\cdot\vk_2)}{k_1^2 k_2^2},  \label{LPTtoF2}\\ 
 F_3(\vk_1,\vk_2,\vk_3) &= \frac{1}{6} \left[ \Gamma_3(\vk_1,\vk_2,\vk_3) + \frac{(\vk_{123}\cdot\vk_1)(\vk_{123}\cdot\vk_2)(\vk_{123}\cdot\vk_3)}{k_1^2 k_2^2 k_3^2} \right] \nonumber\\
                 &\quad  +  \frac{1}{14}\left[\frac{(\vk_{123} \cdot \vk_{12})(\vk_{123} \cdot \vk_{3})}{k_{12}^2 k_3^2}\Gamma_2(\vk_1,\vk_2)+ \text{(cyclic)}\right].
\end{align}
The computation of the $G_n$ kernels requires more work. The peculiar velocity field is given by $\vv = a \dot{\Ps}$, or its divergence
 \begin{equation}\label{divvToPsi}
 J(\vq,t) \frac{1}{a} \frac{\partial\,}{\partial x^i }v^i(\vx,t) = \dot{\Psi}_{i,i} + \Psi_{j,j}\dot{\Psi}_{i,i} - \Psi_{i,j}\dot{\Psi}_{i,j} 
 + \frac{1}{2}\epsilon_{ikp}\epsilon_{jqr} \Psi_{k,q} \Psi_{p,r}\dot{\Psi}_{i,j},
 \end{equation}
 where a colon means derivative with respect to Lagrangian coordinates, and
 $J=\det (J_{ij})$ with $J_{ij}= \partial x_i/\partial q^j$ is the Jacobian matrix of the coordinate transformation of eq.~\eqref{qTox}. In order to arrive to eq.~\eqref{divvToPsi} we use $\partial_{\vx^i}= (J^{-1})_{ji}\partial_{\vq^j} $ and the relation $J\,(J^{-1})_{ji} =  \delta_{ij} + (\delta_{ij}\delta_{ab} - \delta_{ia}\delta_{jb})\Psi_{a,b} + \frac{1}{2}\epsilon_{ikp}\epsilon_{jqr} \Psi_{k,q} \Psi_{p,r}$. With these identities at hand, we compute
%
\begin{align} \label{tfpsid}
\theta_{cb}(\vk) &= -\frac{1}{a H f_0 }\int d^3 x\; e^{-i\vk \cdot \vx}  \frac{\partial v_i}{\partial x^i} 
               =  -\frac{1}{H f_0 }\int d^3 q\; e^{-i\vk \cdot \vq} e^{-i\vk\cdot \Ps(\vq,t)} J(\vq,t) \frac{1}{a} \frac{\partial v_i}{\partial x^i} \nonumber\\
              &= -\frac{1}{H f_0 } \sum_{m=0}^{\infty} \frac{(-i)^m}{m!}\int d^3 q\; e^{-i\vk \cdot \vq} \big(\vk \cdot \Ps\big)^m \Bigg[    \dot{\Psi}_{i,i}    +  (\delta_{ij}\delta_{ab} - \delta_{ia}\delta_{jb})  \Psi_{a,b}\dot{\Psi}_{i,j}    \nonumber\\
              &\quad \qquad +  \frac{1}{2}\epsilon_{ikp}\epsilon_{jqr}    \Psi_{k,q} \Psi_{p,r}\dot{\Psi}_{i,j} \Bigg], 
\end{align}
where we use $d^3x= J(\vq,t)d^3q$ in the second equality, and Taylor-expand the exponential $e^{-i\vk\cdot \Ps(\vq,t)}$ together with eq.~\eqref{divvToPsi} in the third equality. Finally, by using eqs.~\eqref{LDDkernels} and \eqref{LDDdkernels} we can relate the $\Gamma_n$ and $\Gamma^f_n$ LPT kernels to the $G_n$ EPT kernels, to obtain the following expression for $G_2$
\begin{align}
G_2(\vk_1,\vk_2)  &= \frac{3}{7} \Gamma^f_2(\vk_1,\vk_2) + \frac{(\vk_1\cdot\vk_2)^2}{k_1^2 k_2^2} \frac{f_1+f_2}{2f_0} + \frac{1}{2}\frac{\vk_1\cdot\vk_2}{k_1 k_2}  \left( \frac{k_2}{k_1}\frac{f_2}{f_0}+\frac{k_1}{k_2}\frac{f_1}{f_0} \right), \label{LPTtoG2}
\end{align}
where $f_{1} = f(k_{1})$, $f_{2} = f(k_{2})$. 
The third-order kernel $G_3(\vk_1,\vk_2,\vk_3)$ is much more cumbersome and its expression is not very illuminating; while it can be found in eq.~(C.22) of \cite{Aviles:2020wme}. However, it simplifies considerably when it is symmetrized over its three arguments and evaluated at the wave-vectors $\vk_1=\vk$, $\vk_3=-\vk_2=\vp$, which correspond to the configuration appearing in the one-loop computations. By doing this, one arrives at
\begin{align}\label{G3}
& G_3(\vk,-\vp,\vp) =  
   \frac{1}{2} \Gamma^{f}_3(\vk,-\vp,\vp)   +  \frac{2}{7} \frac{\vk \cdot \vp}{p^2} \Gamma^{f}_2(\vk,-\vp)   + \frac{1}{7}  \frac{f(p)}{f_0}  \Gamma_2(\vk,-\vp) \frac{\vk \cdot (\vk-\vp)}{|\vk-\vp|^2} \nonumber\\
&\quad - \frac{1}{6} \frac{(\vk \cdot \vp)^2}{p^4} \frac{f(k)}{f_0}     -  \frac{1}{7}   \left[ 2 \Gamma_2^f(\vk,-\vp) + \Gamma_2(\vk,-\vp) \frac{f(p)}{f_0} \right]\left[1 -\frac{(\vp \cdot (\vk-\vp))^2}{p^2|\vk-\vp|^2}  \right].
\end{align}
The $F_3$ kernel for the same configuration reduces to
\begin{align}\label{F3}
 F_3(\vk,-\vp,\vp)  &= \frac{1}{6}  \Gamma_3(\vk,-\vp,\vp)  +  \frac{1}{7} \frac{\vk\cdot(\vk-\vp) \, \vk \cdot \vp}{p^2|\vk-\vp|^2} \Gamma_2(\vk,-\vp)  - \frac{1}{6} \frac{(\vk \cdot \vp)^2}{p^4}. 
\end{align}
It is easy to show that $F_3$ and $G_3$ given by eqs.~\eqref{F3} and \eqref{G3} reduce to the standard kernels for EdS evolution with the use of 
\begin{align}
\Gamma_2^\text{EdS}(\vk,-\vp) = \Gamma_2^{f,\,\text{EdS}}(\vk,-\vp) &= 1-\frac{(\vk\cdot \vp)^2}{k^2 p^2}, \\
\Gamma_3^\text{EdS}(\vk,-\vp,\vp) = \Gamma_2^{f,\,\text{EdS}}(\vk,-\vp,\vp) &= \frac{5}{21} \frac{k^2}{|\vk-\vp|^2} \left( 1-\frac{(\vk\cdot \vp)^2}{k^2 p^2}\right)^2  + \, (\vp \rightarrow -\vp).  
\end{align}

\subsection{Second order kernels and the tree-level bispectrum}

We now take a closer look to the second order kernels, which are explicitly given in eqs.~\eqref{F2_kernelapp} and \eqref{G2_kernelapp}. It is instructive to write them as 
 \begin{align}
 F_2(k_1,k_2,x) &= \left(\frac{2}{3} + \frac{3\mA - \mB}{14}\right)\mathcal{L}_0(x) + \frac{1}{2}\left(\frac{k_2}{k_1} + \frac{k_1}{k_2} \right) \mathcal{L}_1(x)  + \left( \frac{1}{3} - \frac{\mB}{7} \right)\mathcal{L}_2(x), \label{F2_kernelleg} \\
 G_2(k_1,k_2,x) &= \left( \frac{3(3\mA-\mB)(f_1+f_2) + 3 (3\dot{\mA}-\dot{\mB})/H + 7(f_1+f_2)}{42 f_0} \right)\mathcal{L}_0(x)  \nonumber\\
 &\quad + \frac{1}{2}\left( \frac{f_2}{f_0}\frac{k_2}{k_1}+ \frac{f_1}{f_0}\frac{k_1}{k_2} \right) \mathcal{L}_1(x)  
  + \left( \frac{(3 \mB - 7)(f_1+f_2)+ \dot{\mB}/H}{21 f_0}\right)\mathcal{L}_2(x), \label{G2_kernelleg}
 \end{align}
 where $x\equiv \hat{k}_1\cdot\hat{k}_2$ and $\mathcal{L}_L(x)$ denotes the Legendre polynomial of degree $L$.  The functions $\mA$ and $\mB$ depend on the wave-vector magnitudes $k_1$, $k_2$ and the cosine of their corresponding opening angle, $x$. 
   \revised{These functions show up because of the scale- and time-dependence of the growth rate $f$; or more generally, because the relation $f^2 = \Omega_m(t)$ does not hold. Therefore, one cannot solve analytically the differential equations for the kernels, as it happens in the simplest case of EdS where both the growth rate and the matter abundance are unity \cite{Bernardeau:2001qr}. As such, these functions are}
 solutions to second-order linear differential equations, which can be found in \cite{Aviles:2020cax} [eqs.~(3.9)-(3.11)]; and also in eqs.~\eqref{DAeveq} and \eqref{DBeveq} in appendix \ref{app:2EPTKernels}. In $\Lambda$CDM, the functions $\A$ and $\B$ are scale-independent and evolve very slowly with time, becoming exactly unity for EdS. 
 However, despite its appearance, notice that the above equations are not complete Legendre decompositions since $\mA$ and $\mB$ are themselves functions of $x$. 
For example, the dipole coefficients ($L=1$) are
\begin{align}
 F_2^{L=1}(k_1,k_2) &=   \frac{1}{2}\left(\frac{k_2}{k_1} + \frac{k_1}{k_2} \right) + [\Delta F_2(k_1,k_2,x)]^{L=1}, \\
 G_2^{L=1}(k_1,k_2) &=   \frac{1}{2}\left(\frac{k_2}{k_1}\frac{f(k_2)}{f_0} + \frac{k_1}{k_2}\frac{f(k_1)}{f_0} \right) + [\Delta G_2(k_1,k_2,x)]^{L=1}, 
\end{align}
where the first pieces in the above equations are the so-called gradient terms, while the pieces on squared brackets are given by
\begin{align}
[\Delta F_2(k_1,k_2,x)]^{L=1} &= \frac{3}{2}\int_{-1}^{1} d x \left[F_2(k_1,k_2,x) - \frac{x}{2}\left(\frac{k_2}{k_1} + \frac{k_1}{k_2} \right)  \right] \mathcal{L}_1(x), \label{DF2L1}\\
[\Delta G_2(k_1,k_2,x)]^{L=1} &= \frac{3}{2} \int_{-1}^{1} d x \left[ G_2(k_1,k_2,x) - \frac{x}{2}\left( \frac{k_2}{k_1}\frac{f(k_2)}{f_0} + \frac{k_1}{k_2}\frac{f(k_1)}{f_0} \right)   \right] \mathcal{L}_1(x). \label{DG2L1}
\end{align}
Actually, it turns out that $\Delta G_2$ and $\Delta F_2$ are negligible contributions, because
\begin{align}
 [\Delta F_2(k_1,k_2,x)]^{L=1} &\ll \frac{1}{2}\left(\frac{k_2}{k_1} + \frac{k_1}{k_2} \right), \label{compF2L1}\\
 [\Delta G_2(k_1,k_2,x)]^{L=1} &\ll \frac{1}{2}\left(\frac{k_2}{k_1}\frac{f(k_2)}{f_0} + \frac{k_1}{k_2}\frac{f(k_1)}{f_0} \right).  \label{compG2L1} 
\end{align}

 \begin{figure}
 	\begin{center}
 	\includegraphics[width=5.5 in]{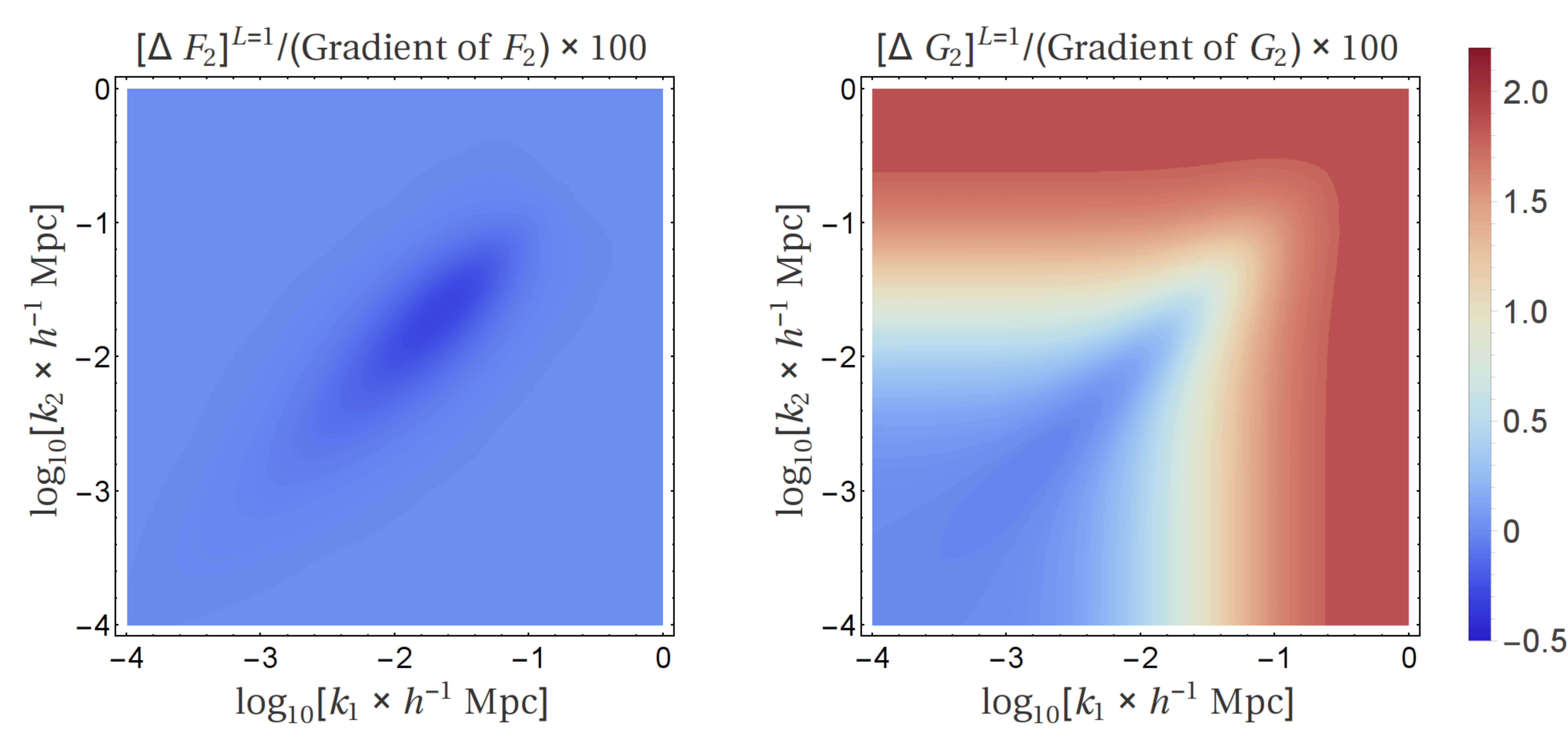}
 	\caption{Ratios of dipole terms $[\Delta F_2]^{L=1}$ and $[\Delta G_2]^{L=1}$, defined in eqs.~\eqref{DF2L1} and \eqref{DG2L1}, over the gradient terms of $F_2$ and $G_2$. These plots show numerically that eqs.~\eqref{compF2L1} and \eqref{compG2L1} hold true. We use $M_\nu=0.4\,\text{eV}$ for which the ratios are the largest, reaching at most  $2\%$ for the $G_2$ kernel within the shown $k$ range. The evaluation redshift is $z=0.5$.  
 	\label{fig:ratiosK}}
 	\end{center}
 \end{figure}
In figure \ref{fig:ratiosK}, we show density plots of the ratios of the LHS to the RHS sides in above inequalities, showing that these inequalities indeed hold.
In conclusion, the dipoles for the $F_2$ and $G_2$ kernels are approximated, to high accuracy, by
\begin{align}
 F_2^{L=1}(k_1,k_2) &=   \frac{1}{2}\left(\frac{k_2}{k_1} + \frac{k_1}{k_2} \right) , \label{F2gradients}\\
 G_2^{L=1}(k_1,k_2) &=   \frac{1}{2}\left(\frac{k_2}{k_1}\frac{f(k_2)}{f_0} + \frac{k_1}{k_2}\frac{f(k_1)}{f_0} \right).  \label{G2gradients}
\end{align}
Moreover, when both momenta are larger than the free streaming wavenumber ($k_1,k_2 > k_\text{FS} \simeq 0.1 \hmpci$), the last expression reduces to
\begin{equation}
 G_2^{L=1}(k_1,k_2) \simeq \frac{1}{2}\left(\frac{k_2}{k_1} + \frac{k_1}{k_2} \right) \left(1-\frac{3}{5}f_\nu \right), 
\end{equation}
where we used the approximation \cite{Hu:1997vi,Wong:2008ws} (see also figure \ref{fig:Dp_fk})
\begin{equation}
 f(k > k_\text{FS}) = \left(1-\frac{3}{5}f_\nu \right) f_0.
\end{equation}
As a result, we obtain a dipole piece in agreement with \cite{Kamalinejad:2020izi}, namely  
\begin{align}
 [F_2^\text{EdS}(\vk_1,\vk_2) - F_2(\vk_1,\vk_2)]^{L=1} &= 0, \label{DF2ELL1} \\
 [G_2^\text{EdS}(\vk_1,\vk_2) - G_2(\vk_1,\vk_2)]^{L=1} &= -\frac{3}{10} \left(\frac{k_2}{k_1} + \frac{k_1}{k_2} \right)f_\nu, \label{DG2ELL1}
\end{align}
which as we stated above, is valid for $k_1,k_2 > k_\text{FS}$. It is worth noticing that this result was obtained in \cite{Kamalinejad:2020izi} using the kernels of \cite{Wong:2008ws}, which are different from ours \revised{since that PT assumes the neutrino density field is linear}; in particular, the gradient terms do not take the form of eqs.~\eqref{gradientF2} and \eqref{gradientG2} below. 
\revised{Despite these differences, the $G_2$ and $F_2$ kernels of \cite{Wong:2008ws} coincide with ours at two important limits. First, at $k \ll k_\text{FS}$ since in this limit neutrino densities are linear and behave as CDM; and second, at small scales, $k\gg k_\text{FS}$, where the neutrino overdensities are essentially zero regardless if these are considered linear or not. Because of the latter, the results in eqs. \eqref{DF2ELL1} and \eqref{DG2ELL1} coincide in both approaches.} 

On small scales within the free-streaming scale, \cite{Kamalinejad:2020izi} applied the expression of \cite{Wong:2008ws} for the linear growth rate, and derived leading-order in $f_{\nu}$ corrections to the $F_2$ and $G_2$ kernels to construct the tree-level bispectrum. Importantly, they found a change to the dipole of the $G_2$ kernel, which dipole is at the pre-cyclic level protected from any changes due to biasing. This is a signature of how neutrinos truly modify velocities in a unique manner.


The reader may also notice that the gradient terms on the EPT kernels are fixed by the advection of density and velocity fields due to the Lagrangian displacement. In the case of the $F_2$ gradient term, it arises from expanding $\delta(\vx+\Ps)$ to lowest order, which results in 
\begin{equation}
\delta^{(1)}(\vx+ \Ps) -\delta^{(1)}(\vx) =\Psi^{(1)}_i(\vx) \partial_i \delta^{(1)}(\vx) = \partial_i\big[\nabla^{-2}\delta^{(1)}(\vx)\big] \partial_i \delta^{(1)}(\vx). 
\end{equation}
The previous expression is part of the second order perturbation,  $\delta^{(2)}(\vx)$, and is the same contribution that one gets in the massless neutrino $\Lambda$CDM model. At this level the relation with the gradient terms of $F_2(\vk_1,\vk_2)$ is not evident, but once we move to Fourier space and symmetrize over the momenta $\vk_1$ and $\vk_2$, the previous expansion becomes
\begin{equation} \label{gradientF2}
\frac{x}{2} \left(\frac{k_2}{k_1} + \frac{k_1}{k_2}\right),
\end{equation}
which clearly corresponds to the gradients terms in dipole coefficient of eq.~\eqref{F2gradients}.
(Do not confuse the Eulerian coordinate $\vx$ with the angle's cosine $x$.) On the other hand, for the $G_2$ one cannot take the $F_2$ result unmodified, because of the relation between the linear overdensities and the velocities (see eq.~\eqref{lineartheta}). Therefore, a similar expansion, but for $\theta^{(1)}(\vx+ \Ps) -\theta^{(1)}(\vx)$, in Fourier space now leads to
\begin{equation} \label{gradientG2}
\frac{x}{2} \left(\frac{f(k_2)}{f_0}\frac{k_2}{k_1} + \frac{f(k_1)}{f_0}\frac{k_1}{k_2}\right),  
\end{equation}
which again is the gradient term in eq.~\eqref{G2gradients} times $\mathcal{L}_1$. 
\revised{As a result, the advection of matter fields differs from the massless neutrino case because of the scale-dependent growth of the linear velocity divergence.}

We end this section by discussing other approaches in the literature.
In the seminal article \cite{Saito:2008bp}, the nonlinear $cb$ fluctuations are obtained neglecting the neutrino density fluctuation and using the EdS kernels, but with the linear power spectrum of the $cb$ fluid as input for the loop computations. 
Another recurrent approach is to approximate $\delta_\nu = \delta^{(1)}_\nu$, obtained from a Boltzmann code, and use the linear fluctuation as an external source to compute the non-linear $cb$ overdensities \cite{Saito:2009ah,Wong:2008ws,Wright:2017dkw}. However, the latter approach violates momentum conservation, as proved in \cite{Blas:2014hya}; in particular, one does not obtain the relation $(P_\text{NL}-P_\text{L})/P_\text{L} \propto k^2$  
at large scales. Instead, the prescription of \cite{Blas:2014hya} evolves the non-linear neutrino density fields by truncating the Boltzmann hierarchy at the Euler equation, and approximating the second moment of the phase-space distribution function to be proportional to an effective sound speed times the density contrast. This \revised{fluid} approach finds justification in the work of \cite{Shoji:2009gg,Shoji:2010hm,Castorina:2015bma} (see also appendix C of \cite{Aviles:2015osc}). 
\revised{In more recent works \cite{Senatore:2017hyk,deBelsunce:2018xtd},} the authors use non-linear perturbations around a Fermi-Dirac massive neutrino distribution and solve the coupled Boltzmann and CDM density field equations iteratively, by expanding in powers of $f_\nu$ and keeping only the linear order.
In addition, it is worth mentioning a paper published very recently in the subject \cite{Garny:2020ilv}, where
neutrino's field fluctuations are described by a hybrid Boltzmann-two fluid model, which then are used to compute the matter real-space power spectrum up to two loops. \revised{Finally, in a similar manner as the renormalization group time flow scheme used in \cite{Lesgourgues:2009am,Upadhye:2013ndm}, or the work in \cite{Levi:2016tlf} which treats $f_\nu$ as an expansion parameter, we improve the description of neutrino evolution by using the non-linear quantity $\delta_{cb} (\delta_\nu^{(1)}/\delta_{cb}^{(1)})$ as a proxy for $\delta_\nu$, instead of consider it a linear quantity or simply neglect it. However, in our approach, we further find exact kernels, up to a few functions that are solutions to linear second-order ordinary differential equations, and which nonetheless are very close to unity. The simple form of our kernels allows us to draw rapid conclusions of the theory and to develop efficient algorithms for the computation of loop corrections; as we do in section \ref{sect:FFTlog}. To the best of our knowledge, the approach presented here is the only one that explicitly recovers the gradient terms in eqs.~\eqref{gradientF2} and \eqref{gradientG2} that are fixed by the advection of matter fields.}

\end{section}
   
\begin{section}{1-dimensional spectra}\label{sect:1dimSpectra}

A direct consequence of the results presented so far is the one-loop auto and cross power spectra for the dentsity and velocity fields. The resulting expressions for the three possible combinations are
\begin{align}
P^\text{1-loop}_{cb,\delta\delta}(k) &=  P^L_{cb,\delta\delta}(k)   + 2 \int_{\vp} \big[F_2(\vp,\vk-\vp)\big]^2 P_L(p) P_L(|\vk-\vp|) \nonumber\\
&\quad + 6 \int_{\vp} F_3(\vk,-\vp,\vp) P_L(k) P_L(p), \label{Pddloop} \\
P^\text{1-loop}_{cb,\delta\theta}(k) &=  P^L_{cb,\delta\theta}(k)   + 2 \int_{\vp} F_2(\vp,\vk-\vp)G_2(\vp,\vk-\vp) P_L(p) P_L(|\vk-\vp|) \nonumber\\
&\quad + 3 \int_{\vp} \left[F_3(\vk,-\vp,\vp)\frac{f(k)}{f_0} + G_3(\vk,-\vp,\vp)\right] P_L(k) P_L(p),  \label{Pdtloop} \\
P^\text{1-loop}_{cb,\theta\theta}(k) &=  P^L_{cb,\theta\theta}(k)   + 2 \int_{\vp} \big[G_2(\vp,\vk-\vp)\big]^2 P_L(p) P_L(|\vk-\vp|) \nonumber\\
&\quad + 6 \int_{\vp} G_3(\vk,-\vp,\vp) \frac{f(k)}{f_0} P_L(k)  P_L(p), \label{Pttloop}
\end{align}
where, as before, we have adopted the compact integral notation $\int_{\vp} =\int d^3p/(2\pi)^{3}$, and the linear contributions $P^L_{cb,\dots}$ are given by
\begin{align}
  P_L(k) \equiv P^L_{cb,\delta\delta}(k) = \langle \delta^{(1)}_{cb}(\vk)\delta^{(1)}_{cb}(\vk')\rangle',
\end{align}
and
\begin{align}
P^{L}_{cb,\delta\theta}(k) = \frac{f(k)}{f_0} P_L(k),\qquad P^{L}_{cb,\theta\theta}(k) = \left(\frac{f(k)}{f_0}\right)^2 P_L(k).
\end{align}
Using these results, we depict the one-loop spectra $P_{cb,\delta\delta}(k)$, $P_{cb,\delta\theta}(k)$ and $P_{cb,\theta\theta}(k)$ in figure \ref{fig:Pab_ratios}, at an intermediate redshift of $z=0.5$ for three degenerate sum of neutrino masses ($M_\nu=0.4, 0.2, 0.1$ eV). It is possible to appreciate how the one-loop result departs from the linear spectra at small scales; this happens more noticeable for $\theta$ field than for $\delta$.
\revised{The departure from linear theory is more pronounced for more massive neutrinos, and when velocity fields are involved, because these are affected by the scale dependence of both linear function and the growth rate, whereas the functions $\mathcal{A}$ and $\mathcal{B}$ and their third order counterparts play a subdominant role.} 
Moreover, there is increment in power before one gets the expected suppression with respect to the linear contribution. Actually, in order to observe the differences of the neutrino masses on short scales, the cosmological parameters are the same as the cosmologies used in section \ref{sect:results}, but instead of keeping fixed $\sigma_8$ we anchor all the models to have the same primordial scalar amplitude $A_s = 2.13 \times 10^{-9}$ at a pivotal scale $k=0.05 \, \text{Mpc}^{-1}$. By choosing this normalization, the power spectra for each massive neutrino case coincide at large scales. 

 \begin{figure}
 	\begin{center}
 	\includegraphics[width=6.0 in]{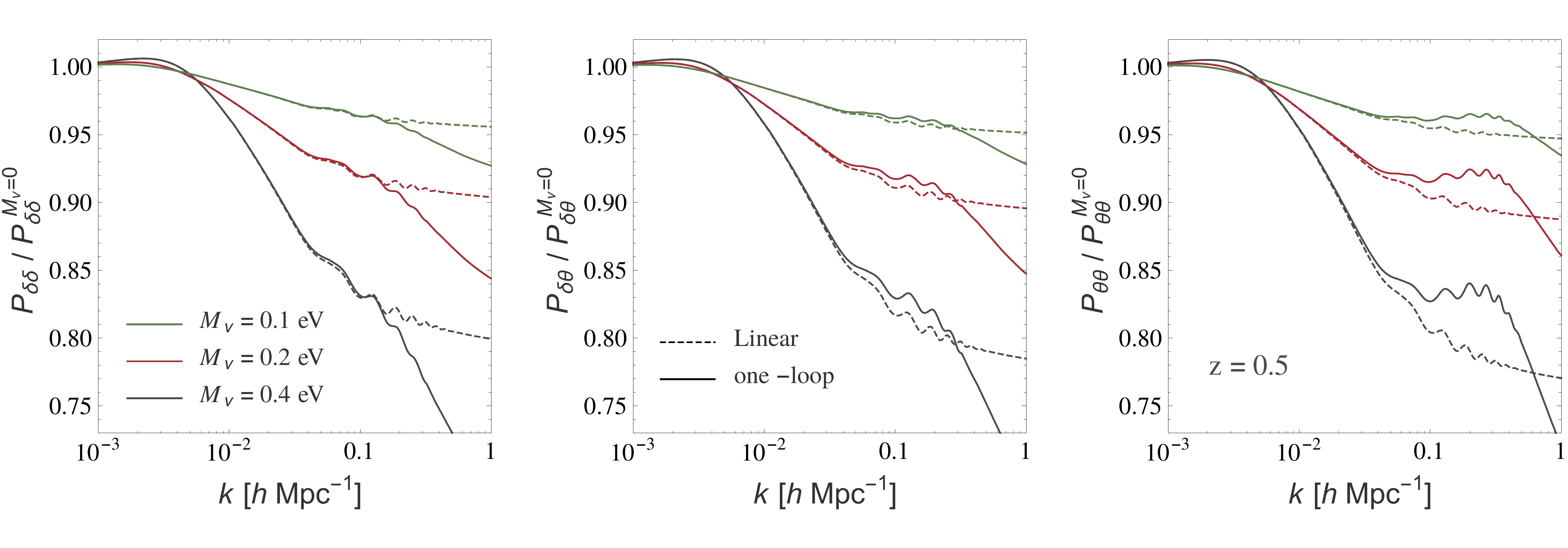}
 	\caption{$cb$ power spectra in the presence of neutrinos with mass $M_\nu = 0.1, 0.2, 0.4\,\text{eV}$ over massless neutrinos $cb$ spectra at redshift $z=0.5$. Left panel shows $P_{\delta\delta}$, middle panel $P_{\delta\theta}$, and right panel $P_{\theta\theta}$. We plot the linear (dashed lines) and one-loop (solid lines) perturbative results computed from eqs.~\eqref{Pddloop}-\eqref{Pttloop}. The spectra are normalized to have the same primordial amplitude $A_s$. 
 	\label{fig:Pab_ratios}}
 	\end{center}
 \end{figure}

\begin{subsection}{Biasing the {\it cb} field}

Large scale bias is primarily sensitive to the clustering of CDM and baryons, with neutrinos playing a subdominant role because their large velocity dispersions prevent them to cluster at the relevant scales and redshifts for halos and galaxies formation. The consequence is that tracers are biased objects of the $cb$ field, and not of the total matter field  \cite{Villaescusa-Navarro:2013pva,Castorina:2013wga,Vagnozzi:2018pwo,Banerjee:2019omr}. Secondary effects, that we do not consider here, arise due to the critical overdensity for collapse dependence on the neutrino mass, affecting the halo abundance and its response to changes in the background density (and hence the biases) \cite{LoVerde:2014pxa,LoVerde:2014rxa}. However, these effects are subdominant and in the following we assume that the biasing of the $cb$ field is universal and constant at large scales.\footnote{Actually \cite{Aviles:2020cax} found a preference for a non-zero small curvature bias for $M_\nu = 0.4 \,\text{eV}$, but consistency with zero for $M_\nu=0.1$ and $0.2\,\text{eV}$.}

We employ the biasing scheme of \cite{McDonald:2009dh}, recently generalized to cosmologies that pose additional scales in \cite{Aviles:2020wme}, which assumes that tracer velocities follow exactly the $cb$ velocities, while tracer density fluctuations $\delta$ are biased by
\begin{align}
\delta(\vx) &= c_\delta \delta_{cb} 
+ \frac{1}{2}c_{\delta^2} \delta_{cb}^2 + c_{s^2} s^2 + \frac{1}{6} c_{\delta^3}\delta_{cb}^3 + \frac{1}{2} c_{\delta s^2} \delta_{cb} s^2 + c_\psi \psi + c_{st} s t + \frac{1}{6} c_{s^3} s^3 \nonumber\\
&\quad + \text{stochastic terms},    
\end{align}
where $s^2(\vx) = s_{ij}s_{ij}$, $s^3(\vx) = s_{ij}s_{jk} s_{ki}$  and 
\begin{equation}
s_{ij}(\vx) = \left( \nabla_i \nabla_j \nabla^{-2} - \frac{1}{3} \delta_{ij} \right) \delta_{cb}(\vx).    
\end{equation}
A useful operator called $\eta(\vx)$ \cite{McDonald:2009dh}, defined here through its Fourier representation
\begin{equation}
    \eta(\vk) = \theta_{cb}(\vk) - \frac{f(k)}{f_0} \delta_{cb}(\vk),
\end{equation}
manifestly vanishes at first order by means of eq.~\eqref{lineartheta}. With $\eta$ in hand, we can construct the following bias operators
\begin{align}
t_{ij}(\vk) &= \left( \frac{k_ik_j}{k^2} - \frac{1}{3} \delta_{ij} \right)  \eta, \\ 
\psi(\vk) &= \eta - \frac{f(k)}{f_0} \left[ \frac{2}{7} s^2 - \frac{4}{21} \delta_{cb}^2\right],
\end{align}
from which we also obtained the third order operator $s t$ as $s t \equiv s_{ij}t_{ij}$. 
It is important to stress that the $\psi$ bias is third order in EdS, but has a second order contribution in the $\Lambda$CDM model with massless neutrinos. Since this second order piece is degenerate with the $s^2$ and $\delta^2$ operators, $\psi$ can be thought to be a pure third order quantity in $\Lambda$CDM \cite{Donath:2020abv}. 
However, in the presence of massive neutrinos, the second order contribution to $\psi$ cannot be absorbed into other operators, but because it is quite small, in the following, we will consider it as a third order operator.

\revised{In the massless neutrino case, the third order bias contributions to the one-loop power spectrum (corresponding to the third order bias operators $st$, $\delta s^2$ and $\psi$) differ by a constant that can be absorbed by the stochastic noise. Hence, they can be collected into a single component given by $b_{3nl}\sigma^2_3(k) P_L(k)$, where $b_{3nl}$ is a non-local bias parameter of third order and $\sigma_3(k)$ takes the form \cite{McDonald:2009dh,Saito:2014qha}}
\begin{equation} \label{sigma23EdS}
 \sigma^{2}_{3}(k) = \frac{105}{16} \ip P_L(p) \left[ S_2(\vp,\vk-\vp)\left(\frac{2}{7}S_2(-\vp,\vk)  -\frac{4}{21} \right) + \frac{8}{63} \right],
\end{equation}
with
\begin{equation}
 S_2(\vk_1,\vk_2) = \frac{(\vk_1\cdot\vk_2)^2}{k_1^2 k_2^2} -\frac{1}{3}.
\end{equation} 
\revised{In contrast, this is no longer possible when additional scales enter into the theory, as in our case the neutrinos' mass scale. Indeed all the contributions have different shapes and one would have to deal with three of them. However, in \cite{Aviles:2020wme} it is shown that adding two bias operators  $\nabla^2\delta$ and $\nabla^2\theta$ suffices to absorb a significant part of the differences in the third order bias contributions which are inherent to a non-EdS evolution. This renormalization-like procedure is not strictly necessary because  one can keep the different biasing shape contributions in the analysis.  However, for fitting purposes it is convenient to deal with just one third-order bias parameter, $b_{3nl}$,  
instead of having three of them that are closely, although not completely, degenerate.}
%
%
Actually, this is the only place where we use the curvature biasing, since we do not use it to construct the scale dependent linear bias or to fit the data. As a matter of fact, \revised{curvature} biases are degenerate with the EFT counterterms that will be introduced later, \revised{and hence terms of the form $k^2 P_L(k)$ do enter into the theory.}

Collecting all the bias contributions and after renormalization, one obtains the tracer spectra to one-loop \cite{Saito:2014qha}
\begin{align}
 P_{\delta\delta}(k) &= b_1^2 P^\text{1-loop}_{cb,\delta\delta}(k)  + 2 b_1 b_2 P_{b_1b_2}(k) + 2 b_1 b_{s^2} P_{b_1b_{s^2}}(k) + b_2^2 P_{b_2^2}(k)  \nonumber\\
                     &\quad + 2 b_2 b_{s^2} P_{b_2 b_{s^2}}(k) + b_{s^2}^2 P_{b_{s^2}^2}(k) + 2 b_1 b_{3nl} \sigma^2_3(k) P^L_{cb,\delta\delta}(k),  \label{PddTNL}\\
 P_{\delta\theta}(k) &=  b_1 P^\text{1-loop}_{cb,\delta\theta}(k) + b_2 P_{b_2,\theta}(k)  + b_{s^2} P_{b_{s^2},\theta}(k) + b_{3nl} \sigma^2_3(k)  P^L_{cb,\delta\theta}(k), \label{PdtTNL}\\
 P_{\theta\theta}(k) &=   P^\text{1-loop}_{cb,\theta\theta}(k),  \label{PttTNL}                 
\end{align}
where the one-loop auto and cross spectra $P^\text{1-loop}_{cb,\delta\delta}(k)$, $P^\text{1-loop}_{cb,\delta\theta}(k)$ and $P^\text{1-loop}_{cb,\theta\theta}(k)$ are given by eqs.~\eqref{Pddloop}, \eqref{Pdtloop} and \eqref{Pttloop}, while the remaining pieces are defined by
\begin{align}
 P_{b_1b_2}(k)     &= \ip F_2(\vp,\vk-\vp) P_L(p)P_L(|\vk-\vp|), \\
 P_{b_1b_{s^2}}(k) &= \ip F_2(\vp,\vk-\vp) S_2(\vp,\vk-\vp) P_L(p)P_L(|\vk-\vp|),  \\
 P_{b_2^2}(k)      &= \frac{1}{2} \ip P_L(p) \big[P_L(|\vk-\vp|) - P_L(p) \big], \\
 P_{b_2 b_{s^2}}(k)&= \frac{1}{2} \ip P_L(p) \left[P_L(|\vk-\vp|)S_2(\vp,\vk-\vp) - \frac{2}{3}P_L(p) \right], \\
 P_{b_{s^2}^2}(k)&= \frac{1}{2} \ip P_L(p) \left[P_L(|\vk-\vp|)[S_2(\vp,\vk-\vp)]^2 - \frac{4}{9}P_L(p) \right] \\
P_{b_2,\theta}(k) &=\ip G_2(\vp,\vk-\vp) P_L(p)P_L(|\vk-\vp|),  \\
P_{b_{s^2},\theta}(k) &=\ip G_2(\vp,\vk-\vp) S_2(\vp,\vk-\vp) P_L(p)P_L(|\vk-\vp|).
\end{align}
Notice that these expressions take the same form as in $\Lambda$CDM, and the only differences rely in the relations between the bare and the renormalized biases \cite{Aviles:2020wme}.

One may think of reducing the number of bias parameters by using co-evolution theory \cite{Chan:2012jj,Baldauf:2012hs,Saito:2014qha}, which mathematically implies that
\begin{equation}\label{coevbiases}
 b_{s^2}= -\frac{4}{7}(b_1-1),\qquad b_{3nl}=\frac{32}{315}(b_1-1).
\end{equation}
These relations assume an EdS evolution, so in the present context of massive neutrinos they are not entirely satisfied. Yet, they only exhibit trends, since its derivation assume that the initial tracer fluctuations are expanded in densities, while tidal and non-local contributions are subsequently generated by the non-linear gravitational evolution. Actually the last statement is not completely true, and indeed a non-vanishing Lagrangian tidal bias has been measured in simulations \cite{Modi:2016dah}. However, we believe that for the statistical level present in our halo catalogues, in principle, the use of the co-evolution relations~\eqref{coevbiases} should be sufficient for parameter estimation.
We will discuss further on this when comparing
the theory to the synthetic data, but before doing so
let us study the effect introduced by RSD.

\end{subsection}

\end{section}

\begin{section}{Redshift-space power spectrum}\label{sect:RSDpk}

Peculiar velocities distort the observed positions of objects in the sky through the Doppler effect. For non-relativistic particles, the effect is on the direction of observation, such that in the distant-observer approximation, the map between the true $\vx$ and the observed $\vs$ positions is
\begin{equation}\label{xTos}
 \vs = \vx + \vu,    
\end{equation}
where the line-of-sight velocity
\begin{equation}
 \vu = \vhn \frac{\vhn \cdot \vv}{a H}, 
\end{equation}
is constructed from the peculiar velocity, $\vv = a \dot{\vx}$, and the angular direction to the sample, $\vhn$. Conservation of number of objects against the coordinate transformation \eqref{xTos} gives 
the redshift-space power spectrum \cite{Scoccimarro:2004tg}
\begin{equation}\label{RSDPS}
  (2\pi)^3\delta_\text{D}(\vk) + P_s(\vk) = \int d^3x e^{-i\vk\cdot \vx} \Big\langle \big(1+\delta(\vx_1)\big)\big(1+\delta(\vx_2)\big)  e^{-i \vk \cdot \Delta \vu}  \Big\rangle , 
\end{equation}
with $\vx = \vx_2 - \vx_1$ and $\Delta \vu = \vu(\vx_2)-\vu(\vx_1)$. By Taylor expanding the exponential inside the correlation of above's equation, the power spectrum becomes \cite{Vlah:2018ygt}
\begin{align} \label{PSmomexp}
   (2\pi)^3 \dD(\vk) + P_s(\vk)
      &= \sum_{m=0}^\infty \frac{(-i)^\tm}{\tm!} (k\mu)^m  \tilde{\Xi}^{(m)}_{\vhn}(\vk),
\end{align}
where $\mu = \hat{\vk} \cdot \vhn$, and $\tilde{\Xi}^{(m)}_{\vhn}$ are the Fourier transforms of the pairwise velocity moments along the line-of-sight,
\begin{align}
 \Xi^{(m)}_{\vhn}(\vx) &\equiv  \langle \big(1+\delta(\vx_1)\big)\big(1+\delta(\vx_2)\big) \big(u(\vx_2)-u(\vx_1) \big)^m \rangle,
\end{align}
 with $u = \vhn  \cdot \vu = |\vu|$.
Notice that the difference of peculiar velocities of tracers located at $\vx_2$ and $\vx_1$ projected along the line-of-sight is given by $(a H) \vhn \cdot \Delta \vu =  \vhn \cdot (\vv(\vx_2) - \vv(\vx_1))$. 
To linear order in perturbation, one truncates the sum in eq.~\eqref{PSmomexp} at $m=2$, to obtain the (linear) Kaiser power spectrum \cite{Kaiser:1987qv}
\begin{equation}\label{PS_K}
    P^K_s(k,\mu) = \left(b_1 + f(k) \mu^2\right)^2 P_L(k), 
\end{equation}
which has the same structure as in the massless neutrinos case, but now the Kaiser boost becomes scale-dependent due to the linear growth rate.

\begin{subsection}{One-loop power spectrum}

We generalize the Kaiser spectrum by cutting the sum in eq.~\eqref{PSmomexp} again at $m=2$, but keeping terms up to one-loop, obtaining the non-linear Kaiser power spectrum
\begin{equation}\label{PS_K_NL}
P^{K,\text{NL}}_s(k,\mu) = P_{\delta\delta}(k) + f_0 \mu^2 P_{\delta\theta}(k) + f_0^2 \mu^4 P_{\theta\theta}(k),   
\end{equation}
with $P_{\delta\delta}$, $P_{\delta\theta}$ and $P_{\theta\theta}$ given by eqs.~\eqref{Pddloop}, \eqref{Pdtloop} and \eqref{Pttloop}.
However, this expression does not contain all one-loop corrections. In fact, the moment $\Xi^{(m)}_{\vhn}$ involves correlations of at least $m$ velocity fields and density fluctuations; therefore, the complete expression for the one-loop EPT power spectrum is obtained by truncating the sum at $m=4$, yielding \cite{Taruya:2010mx,Aviles:2020wme}
\begin{equation}\label{PS_EPT}
P^\text{EPT}_s(k,\mu) = \sum_{m=0}^4 \frac{(-i)^\tm}{\tm!} (k\mu)^m  \tilde{\Xi}^{(m)}_{\vhn}(\vk) = P^{K,\text{NL}}_s(k,\mu) + A^\text{TNS}(k,\mu) + D(k,\mu), 
\end{equation}
while higher order moments are pure two or higher loops contributions. The functions $A^\text{TNS}(k,\mu)$ and $D(k,\mu)$ are \cite{Aviles:2020wme,Taruya:2010mx}
\begin{align}
A^\text{TNS}(k,\mu) &= 2 k \mu f_0 \ip  \frac{\vp\cdot \vhn}{p^2} B_\sigma(\vp,-\vk,\vk-\vp) \, , \label{ATNS} \\
D(k,\mu) &= (k\mu f_0)^2  \ip \Big\{ F(\vp)F(\vk-\vp)  \nonumber\\
      &\quad+   \frac{(\vp\cdot\vhn)^2}{p^4}P_{\theta\theta}^L(p) \big[ P^K_s(|\vk-\vp|, \mu_{\vk-\vp}) -P^K_s(k, \mu) \big] \Big\}, \label{DTNS}
\end{align}
with $\mu_{\vk-\vp}$ the cosine angle between the wave-vector $\vk-\vp$ and the line-of-sight direction $\vhn$, and
\begin{align}\label{defBsigma}
B_\sigma(\vk_1,\vk_2,\vk_3) &= \left\langle \theta(\vk_1)\left[\delta(\vk_2) + f_0 \frac{(\vk_2\cdot\vhn)^2}{k_2^2}\theta(\vk_2) \right]\left[\delta(\vk_3) + f_0 \frac{(\vk_3\cdot\vhn)^2}{k_3^2}\theta(\vk_3) \right]\right\rangle', \\
F(\vp) &= \frac{\vp\cdot\vhn}{p^2}\Big[ P_{\delta\theta}(p) + f_0 \frac{(\vp\cdot\vhn)^2}{p^2} P_{\theta\theta}(p) \Big].
\end{align}
The decomposition in the second equality of eq.~\eqref{PS_EPT} makes sense since $P^{K,\text{NL}}$ is constructed out of correlations of only two fields, either overdensities or velocities, $A$ from three fields, and  $D$ from four fields. 

This model is similar to TNS \cite{Taruya:2010mx}, though the latter does not contain the term $P^K_s(|\vk-\vp|, \mu_{\vk-\vp})$ in the squared brackets of eq.~\eqref{DTNS}, which is necessary to have a well behaved theory, free of both IR and UV divergences once assuming typical linear power spectra \cite{Aviles:2020wme}. Furthermore, the TNS model have a phenomenological Gaussian suppression factor $\exp[k^2\mu^2 f^2 \sigma^2_v]$, providing the damping of the power spectrum along the line-of-sight direction. This term can be interpreted also as a partial factorization of the Fingers-of-God (FoG) and Kaiser boost effects.

Another key aspect to consider is that several of the functions involved in the line-of-sight moments $\Xi^{(m)}_{\vhn}$ contain non-vanishing zero-lag correlators sensible to small-scales physics, so there is a need to add counterterms to these expressions. Specifically, these come from
\begin{align}
  -  \sigma^2_v f_0^2 (\mu^2 + 2 f \mu^4 + f^2 \mu^6) k^2  P_L(k)  \in D(k,\mu), \label{inD}\\
  - \sigma^2_\Psi (1 + 2 f \mu^2 + f^2 \mu^4)k^2  P_L(k)  \in P_s^{K,\text{NL}}(k,\mu), \label{inPK}
\end{align}
with $\sigma^2_\Psi = \int_0^\infty dp \; P_L(p)/(6 \pi^2)$ the 1-dimensional variance of Lagrangian displacements, and  $\sigma^2_v = \int_0^\infty dp \;P^L_{\theta \theta}(p)/(6 \pi^2)$ the velocity variance. Notice that these two correlators are identical for the massless neutrino case, and are simply named  $\sigma^2_v$ in that instance. 
Therefore, the EFT contribution to the power spectrum becomes \cite{Perko:2016puo,Chen:2020fxs}
\begin{equation}\label{EFT_PS}
P^\text{EFT}_s(k,\mu) =  P^\text{EPT}_s(k,\mu) + (\alpha_0 + \alpha_2 \mu^2 + \alpha_4 \mu^4 +\alpha_6 \mu^6) k^2 P_L(k) + P_{\epsilon\epsilon}(k,\mu),     
\end{equation}
where we have neglected the growth rate scale-dependence in eqs.~\eqref{inD} and \eqref{inPK}, since at the scales the counterterms become important, it has reached its small scale limit $f\approx (1-3f_\nu /5 ) f_0$. However, adding the $f(k)$ factors to the above equation can be done without complications.  
Below, we will compute multipoles of the power spectrum up to $\ell=4$, thus the counterterm $\alpha_6$ would be redundant and not considered in this work. One should notice that in real space, EFT contributions are included to model the small scale physics out of the reach of PT, and to tame UV-divergencies (at one-loop present in $P_{13}$). In redshift space the counterterms further model the non-linear relation between overdensities in real and redshift spaces. Being this similar to the approach of some earlier works in the subject, which add to the velocity dispersion variance, $\sigma^2_v$, a phenomenological free parameter, $\sigma^2_\text{FoG}$, to better modelling the FoG.  

With respect to a possible stochastic terms, uncorrelated with long wave-length fluctuations, we use white noise, plus a tilt proportional to $(k\mu)^2$, resulting in the additional power spectra contribution \cite{Perko:2016puo,Chen:2020fxs,Nishimichi:2020tvu,Schmittfull:2020trd,Chen:2020zjt}
\begin{equation}
P_{\epsilon}(k,\mu) =   P_{shot} \big[\alpha^{shot}_0 + \alpha^{shot}_{2} (k\mu)^2 \big],     
\end{equation}
with the standard Poisson process shot-noise as normalization factor $P_{shot} = 1/\bar{n}_X$, where $\bar{n}_X$ is the mean number density of tracers. The departure from a white noise arises because stochasticity is not localized at a single point \cite{McDonald:2009dh,Desjacques:2016bnm} and because the stochastic nature of peculiar velocities at small scales \cite{Chen:2020fxs}. Moreover, stochastic contributions of the form $k^{2n}$ are also necessary for the renormalization of contact terms that typically diverge in Fourier space \cite{McDonald:2006mx,Assassi:2014fva,Aviles:2018thp}.

Finally, and to complete the full bias framework, we need to introduce biasing in functions $A^\text{TNS}$ and $D$. These are obtained by weighting them with the correct power of $b_1$, namely
\begin{align}
A^\text{TNS}(k,\mu;f_0) &\, \longrightarrow \, b_1^3 A^\text{TNS}(k,\mu;f_0/b_1), \\
D(k,\mu;f_0) &\, \longrightarrow \, b_1^4 D(k,\mu;f_0/b_1).
\end{align}
It turns out that the biasing for the $D$ function is exact, because 
$D$ is constructed out of only linear fields. Instead, the function $A^\text{TNS}$ is also biased by the non-linear $\delta^2_{cb}$ and $s^2$ operators. However, their contributions are very small so that we neglect them; see also \cite{Beutler:2013yhm} for using this approximation in the reduction in bias parameters for the TNS model. As a reference, the complete biased expression of $A^\text{TNS}(k,\mu)$, including $b_2$ and $b_{s^2}$ parameters, can be found in appendix A.1 of \cite{Aviles:2020wme}. 

\end{subsection}

\begin{subsection}{IR-resummations}

Long wave-length, bulk displacements of matter tend to degrade the BAO in the power spectrum, because coherent flows stream over a scale settled by (2 times) the 1-dimensional variance of Lagrangian displacements $\sigma^2_\Psi$, which is comparable in size to the BAO peak width. As a result, overdense regions are partially depleted, while underdense regions are partially populated, broadening the acoustic peak \cite{Eisenstein:2006nk,Crocce:2007dt,Tassev:2013rta}. This effect is reasonably well described by LPT, even at its first order, the Zeldovich Approximation (ZA), since bulk flows are captured by the linear displacement fields \cite{Matsubara:2007wj,Carlson:2012bu,Vlah:2015sea,Chen:2020fxs,Chen:2020zjt}. In contrast, in the EPT the convergence is very slow (see e.g.~\cite{Lewandowski:2018ywf}), thus non-perturbative methods are commonly used to account for this effect. Here, we employ IR-resummation methods \cite{Senatore:2014via} to model the degradation of the BAO features. We follow the prescription of \cite{Baldauf:2015xfa,Ivanov:2018gjr,Chudaykin:2020aoj}, that splits the linear power spectrum in a piece that does not contain the BAO (the non-wiggle power spectrum, $P_{nw}$) and a wiggle piece $P_w$;\footnote{We perform this decomposition using the fast sine transform recipe of \cite{Hamann:2010pw}.} such that the real-space linear power spectrum can be written as $P_L = P_{nw} + P_w$. As a result of this splitting, the one-loop IR-resummed EFT redshift-space power spectrum becomes \cite{Ivanov:2018gjr}
\begin{align}\label{PsIR}
P_s^\text{IR}(k,\mu) &= 
 e^{-k^2 \Sigma^2_\text{tot}(k,\mu)} P_s^\text{EFT}(k,\mu) +  \big(1-e^{-k^2 \Sigma^2_\text{tot}(k,\mu)} \big) P_{s,nw}^\text{EFT}(k,\mu) \nonumber\\
 &\quad +  e^{-k^2 \Sigma^2_\text{tot}(k,\mu)} P_w(k) k^2 \Sigma^2_\text{tot}(k,\mu), 
\end{align}
where the wiggle piece $P_s^\text{EFT}(k,\mu)$ is the one-loop power spectrum computed using eq.~\eqref{EFT_PS}. The non-wiggle part, $P_{s,nw}^\text{EFT}(k,\mu)$ is also computed with eq.~\eqref{EFT_PS} but using as input the non-wiggle linear power spectrum $P_{nw}$. Furthermore, the function $\Sigma^2_\text{tot}$ is given by
\begin{equation}\label{Sigma2T}
\Sigma^2_\text{tot}(k,\mu) = \big[1+f \mu^2 \big( 2 + f \big) \big]\Sigma^2 + f^2 \mu^2 (\mu^2-1) \delta\Sigma^2,    
\end{equation}
\revised{where $f$ is the scale-dependent growth rate in the presence of massive neutrinos, and}
\begin{align}
\Sigma^2 &= \frac{1}{6 \pi^2}\int_0^{k_s} dp \,P_{nw}(p) \left[ 1 - j_0\left(p \,\ell_\text{BAO}\right) + 2 j_2 \left(p \,\ell_\text{BAO}\right)\right],\\
\delta\Sigma^2 &= \frac{1}{2 \pi^2}\int_0^{k_s} dp \,P_{nw}(p)  j_2 \left(p \,\ell_\text{BAO}\right),
\end{align}
where $\ell_\text{BAO}\simeq 105 \hmpc$ is the BAO peak scale and $j_n$ are the spherical Bessel functions of degree $n$. The scale $k_s$ separates the long and short modes, whose choice is somewhat arbitrary. However, final results depend weakly on it if $k_s\gtrsim 0.05 \hmpci$. Here we use the value $k_s = 0.2 \hmpci$, following \cite{Chudaykin:2020aoj}. \revised{We use the same standard IR-resummed method as in the prescription of \cite{Ivanov:2018gjr}, with the only difference that the damping \eqref{Sigma2T} is scale-dependent because of the $f(k)$ presence. Notice that we have no formal derivation of this IR-resummation in the presence of massive neutrinos, however, we expect to be at the same level of approximation than in \cite{Ivanov:2018gjr} given that the resummed long-wavelengths behave essentially as in the massless neutrino case.}

The IR-resummed EFT power spectrum of eq.~\eqref{PsIR} is the one we compare the simulations to. More precisely, we take its monopole, quadrupole and hexadecapole multipoles from
\begin{equation}\label{Pells}
P_\ell(k) = \frac{2 \ell + 1}{2} \int_{-1}^{1} d\mu \; P_s^\text{IR}(k,\mu) \mathcal{L}_{\ell}(\mu),    
\end{equation}
where $\mathcal{L}_{\ell}$ are the Legendre polynomial of degree $\ell$.
\end{subsection}

\end{section}

\begin{section}{Numerical results for tracers}\label{sect:results}

 \begin{figure}
 	\begin{center}
 	\includegraphics[width=6 in]{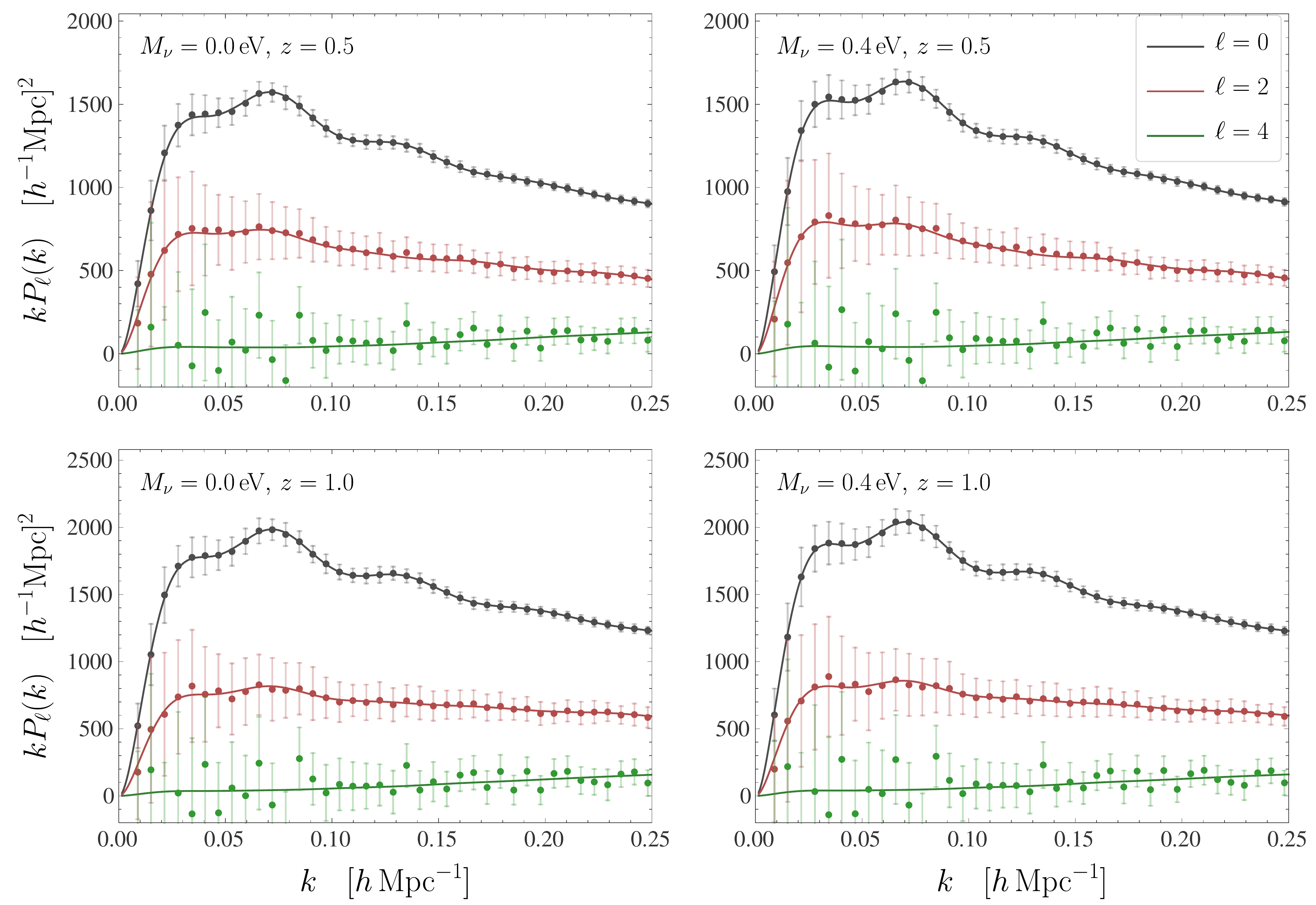}
 	\caption{Comparison of the halo  power spectrum  multipoles obtained from the \textsc{Quijote} simulations and the predictions from the perturbation theory: monopole (black lines), quadrupole (red) and hexadecapole (green). The different panels  show the cases $M_\nu=  0$ and  $0.4$  eV at redshifts $z=0.5$ and $z=1$ (the cases $M_\nu=  0.1$ and  $0.2$  eV give similar results). The  fittings to data are performed up to $k_\text{max}=0.25 \hmpci$ (for comparison of different choices of $k_\text{max}$ see fig.~\ref{fig:kmax}). The analytical monopole results lie within the 1\% from the data points and the quadrupole within the $3\%$. The Poisson noise $P_{shot}$ has been subtracted from the monopole.   
 	\label{fig:multipoles}}
 	\end{center}
 \end{figure}

 \begin{figure}
 	\begin{center}
 	\includegraphics[width=3in]{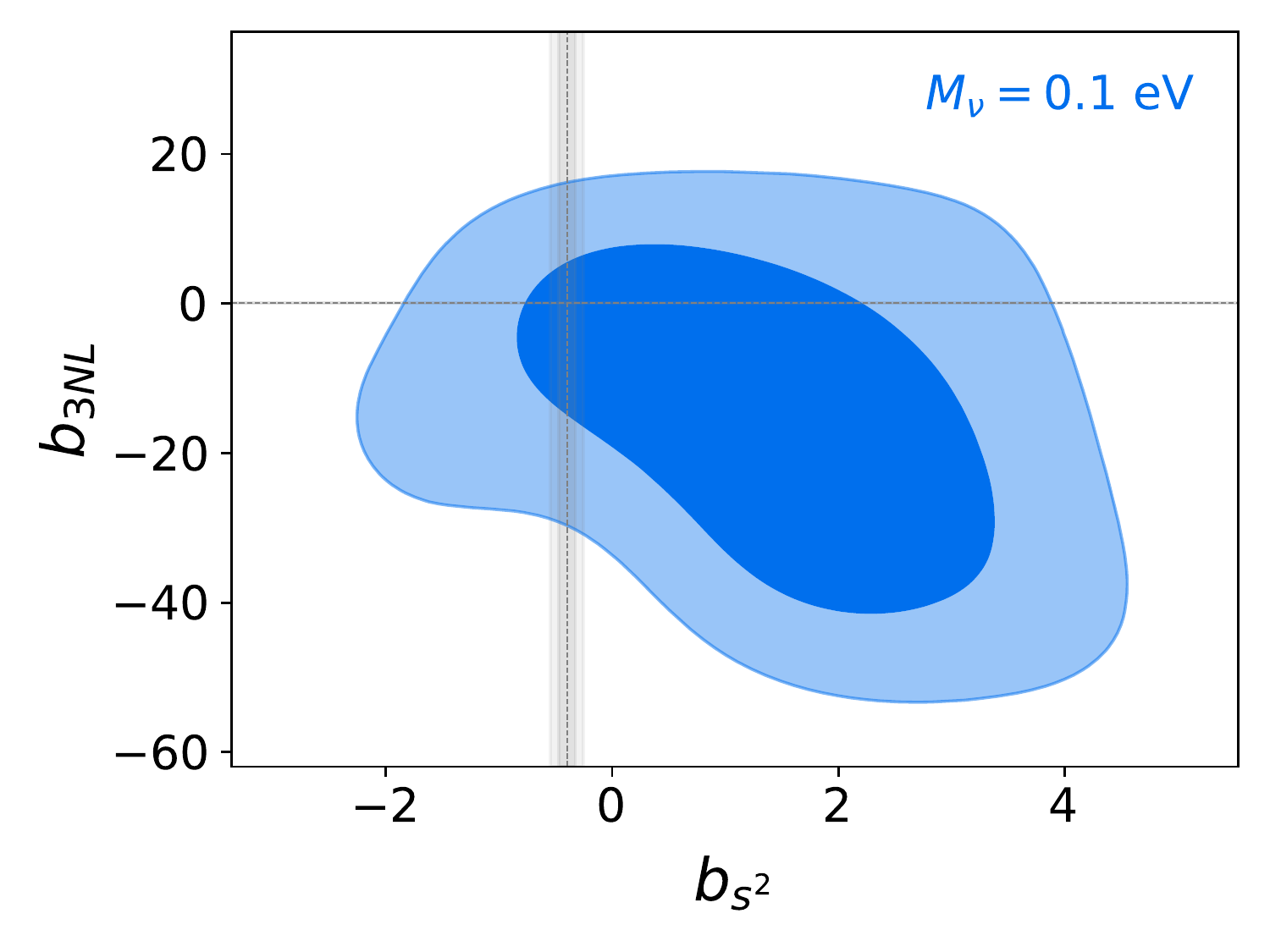}
 	\includegraphics[width=3in]{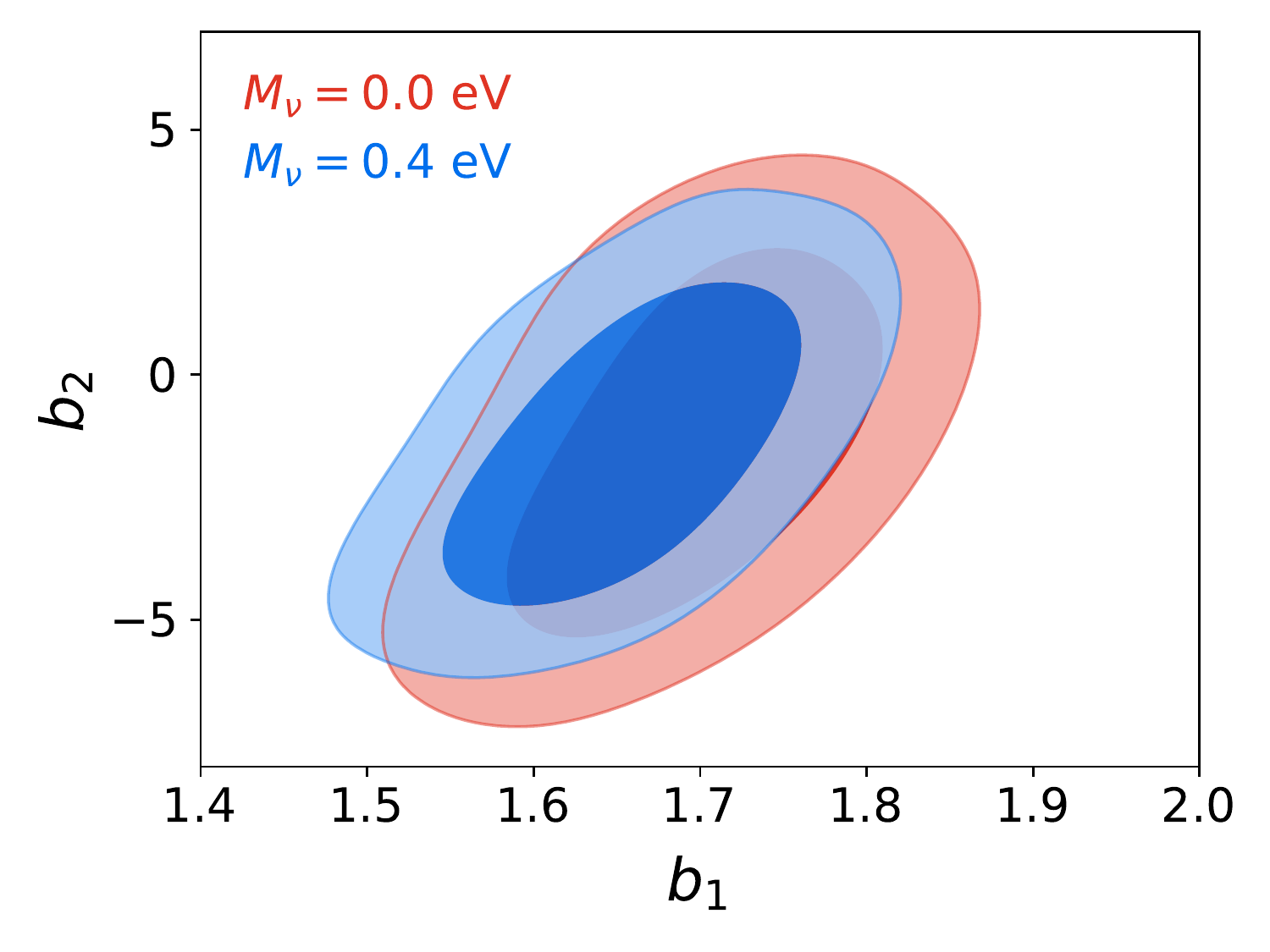}
 	\caption{Posterior contours for combinations of bias parameters in the massive neutrino cosmology with $M_\nu=0.1$ eV at redshift $z=0.5$. In the left panel we show the contour plot at 0.68 and 0.95 C.L. for the subspace $b_{s^2}$-$b_{3nl}$. The gray and vertical bands show the expected values from co-evolution in EdS. In the right panel we show the contour plot for the subspace $b_{1}$-$b_{2}$.  
 	\label{fig:b1b2_coev}}
 	\end{center}
 \end{figure}
 
  \begin{figure}
 	\begin{center}
 	\includegraphics[width=3.0 in]{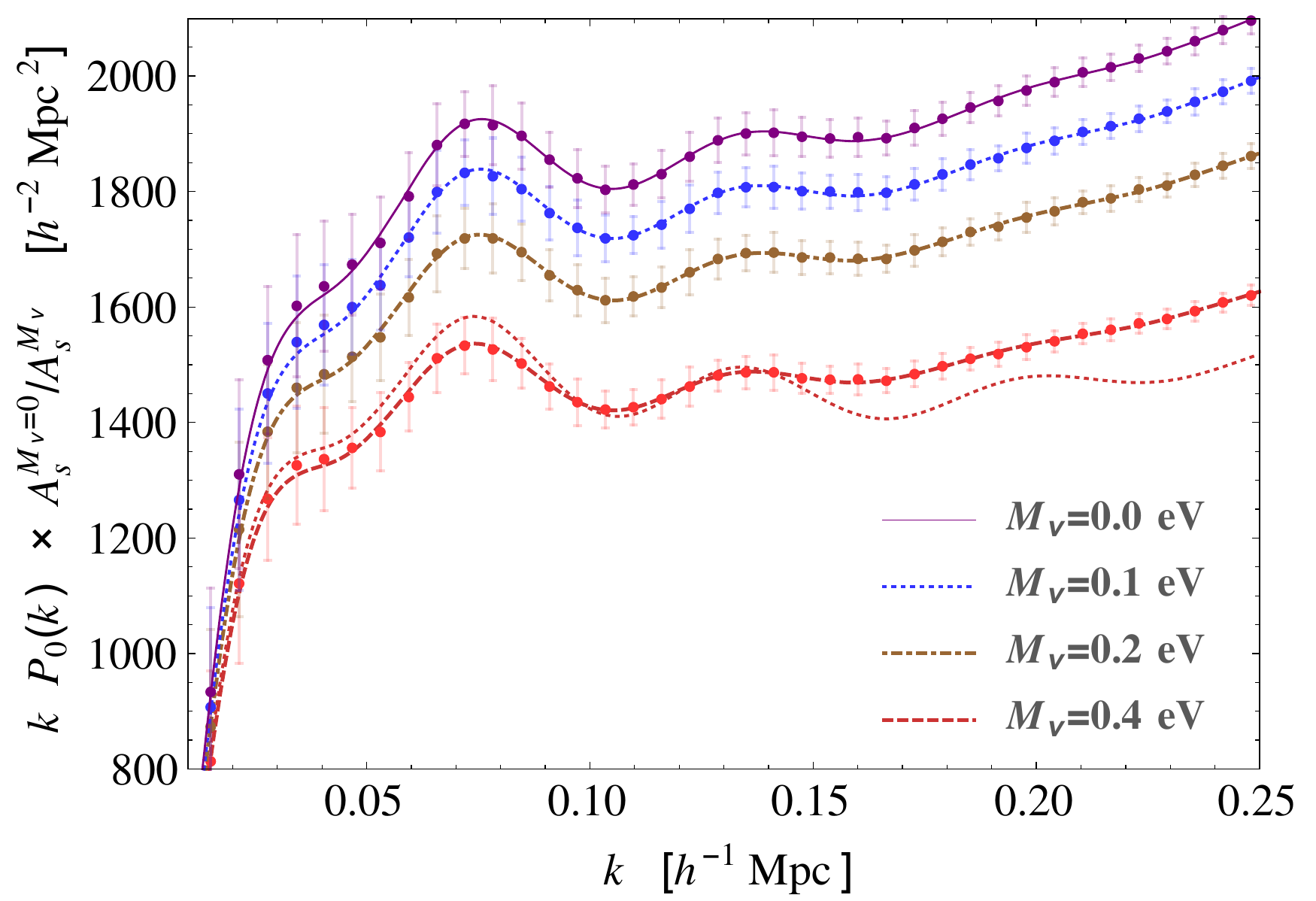}
 	\includegraphics[width=3.0 in]{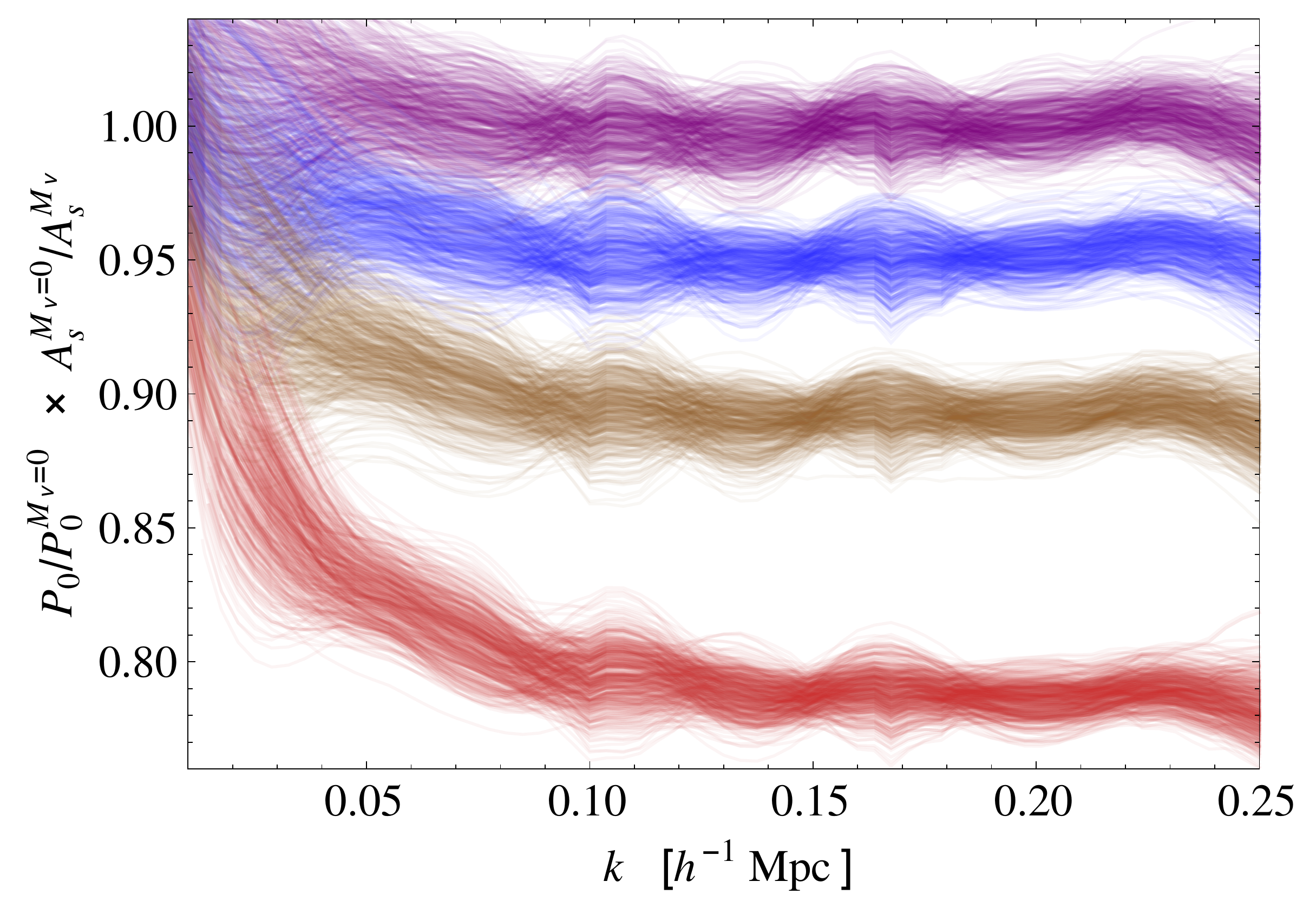}
  	\includegraphics[width=3.0 in]{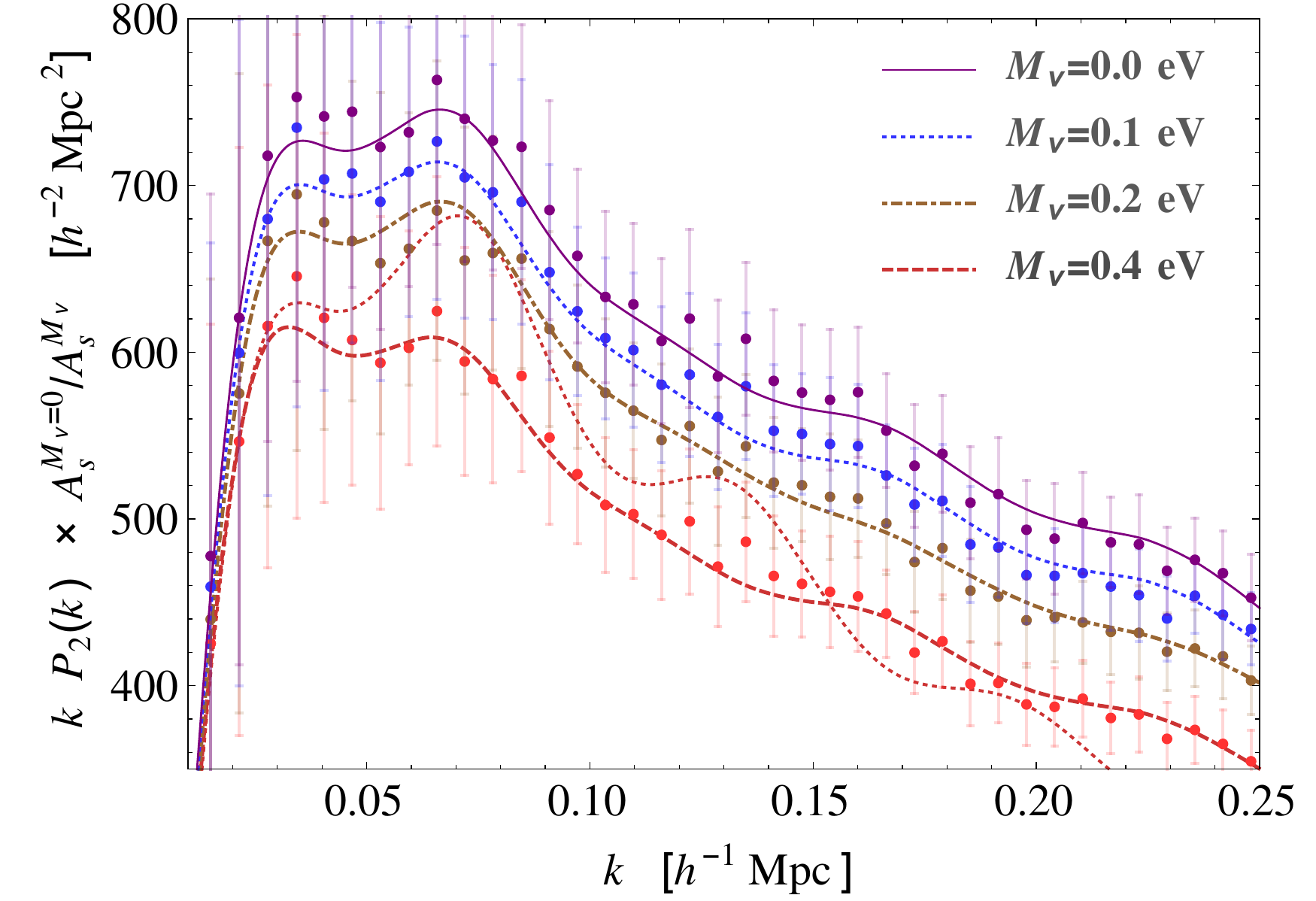}
  	\includegraphics[width=3.0 in]{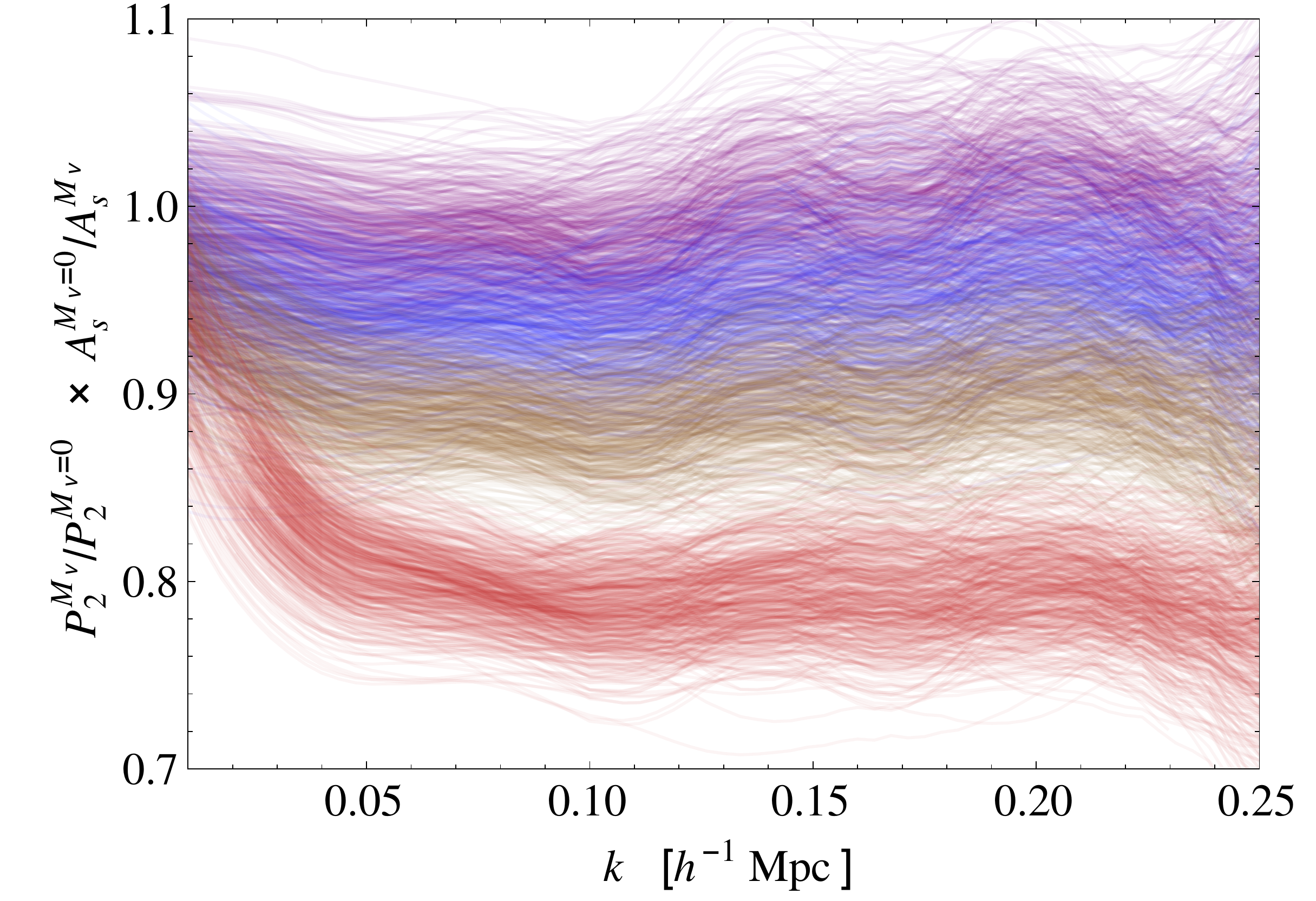}
 	\caption{Power spectrum for tracers: monopole (top panels) and quadrupole (bottom panels) at $z=0.5$ for $M_\nu=0.0\,\text{eV}$ (solid purple line), $M_\nu=0.1\,\text{eV}$ (dotted blue), $M_\nu=0.2\,\text{eV}$ (dot-dashed brown) and 
 	 $M_\nu=0.4\,\text{eV}$ (dashed red) compared to $N$-body simulations data.  
 	For visualization purposes, we have multiplied the results by constant factors 
 	$A_s^{M_\nu=0}/A_s^{M_\nu=0,0.1,0.2,0.4\,\text{eV}}$, such that the 
 	corresponding unbiased multipoles in all models and their linear real space power spectra have the same primordial amplitude, instead of the same $\sigma_8$ normalization. We do this because for fixed $\sigma_8$, the differences among the different massive neutrino cases are located at large scales where the sample variance is large. \revised{The dotted red lines show the linear Kaiser result for the $M_\nu=0.4\,\text{eV}$ case.}
 	The right panels show the ratios over the best fit massless neutrino case, together with a dispersion around the central values by sampling 500 times the parameter posterior distributions.
 	\label{fig:ratios}}
 	\end{center}
 \end{figure}

 \begin{figure}
 	\begin{center}
 	\includegraphics[width=6.0 in]{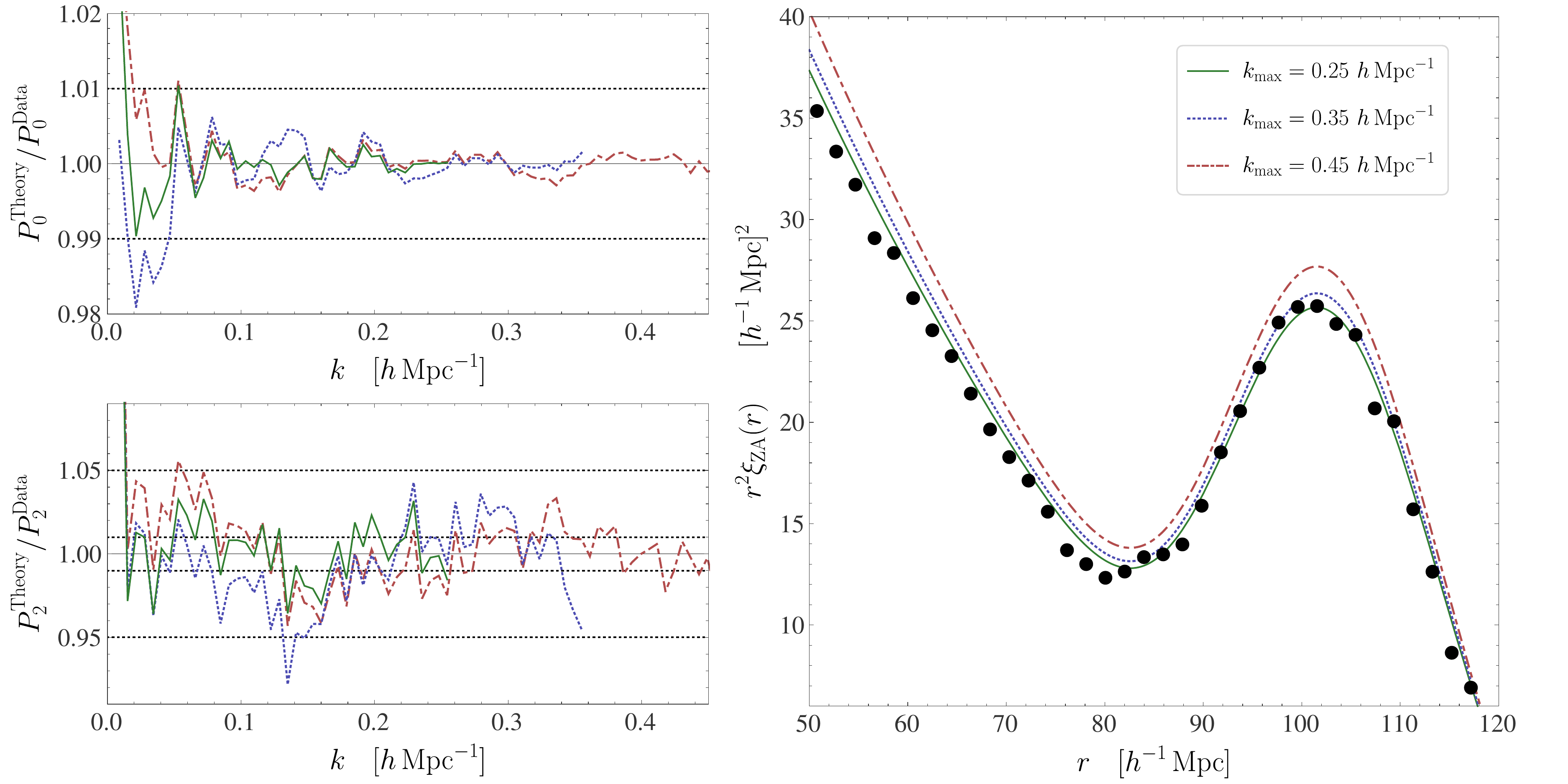}
 	\caption{Comparison when the fitting to data of the halo power spectra is performed up to $k_\text{max}=0.25$, $0.35$ and $0.45 \hmpci$, for sum of neutrino masses $M_\nu=0.1\,\text{eV}$ at redshift $z=0.5$. In the left panels, we show the relative differences of the monopole and quadrupole of the theory and the data from simulations. We observe that the fitting is similar in all cases. In the right panel, we show the Zeldovich Approximation correlation function using the linear biases obtained from the different power spectrum fittings. We observe that as larger is $k_{\rm max}$, the poorer is the correlation function fitting to the data. 
 	\label{fig:kmax}}
 	\end{center}
 \end{figure}

 \begin{figure}
 	\begin{center}
 	\includegraphics[width=4.5in]{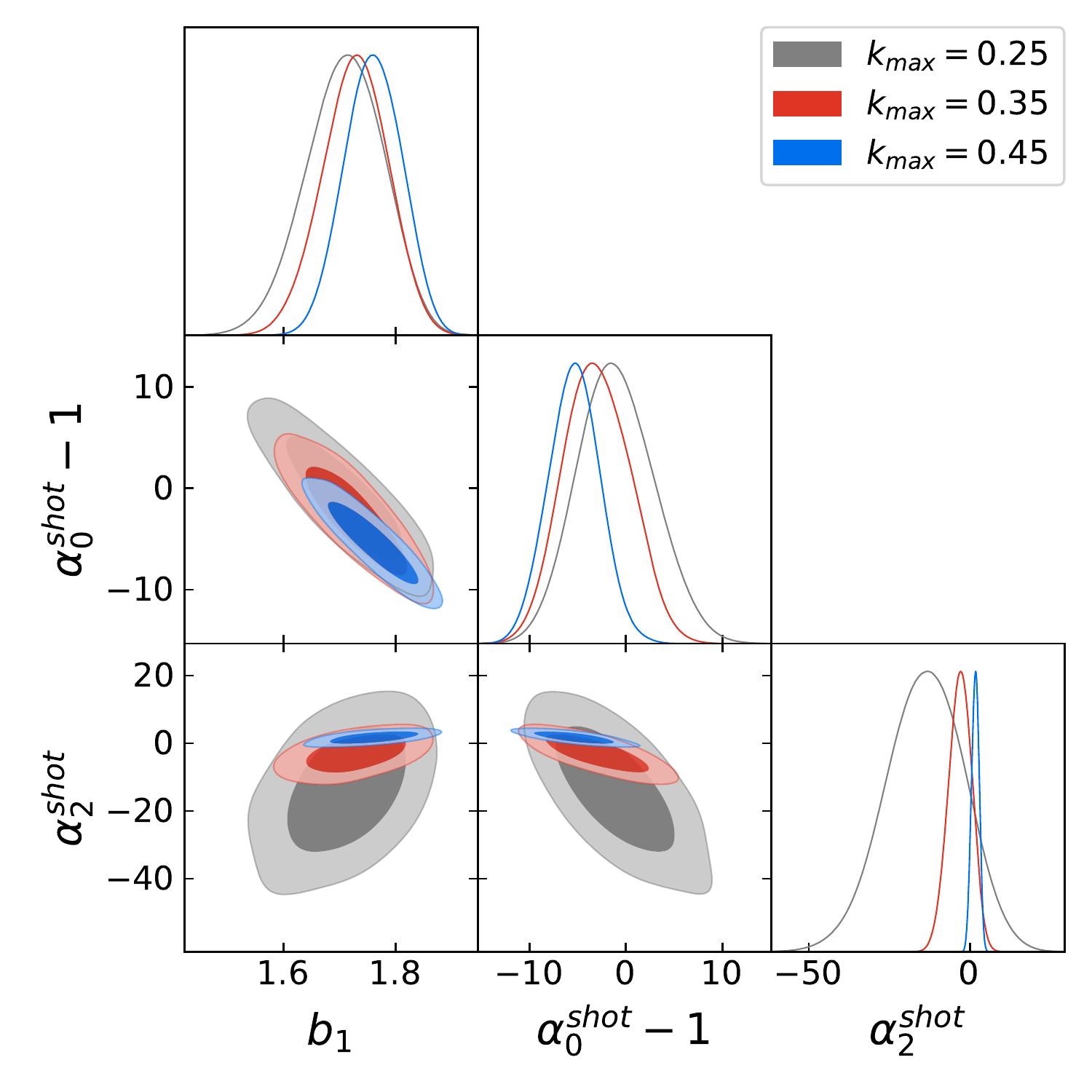}
 	\caption{Posterior contours and 1-dimensional marginalized posterior densities for the linear bias parameter $b_1$ and the shot noise parameter. This plot is for the massive neutrino cosmology with $M_\nu=0.1$ eV at $z=0.5$. The units of $\alpha^{shot}_2$ are $h^{-2} \, \text{Mpc}^2$.  
 	\label{fig:triang}}
 	\end{center}
 \end{figure}

We are now in position to compare our analytical predictions to the halo redshift space power spectrum multipoles obtained from $N$-body simulations.   To this end we use the \textsc{Quijote} suite of simulations, whose fiducial cosmology is $\{ \Omega_m =0.3175, \Omega_b=0.049, h=0.6711, n_s=0.9624, \sigma_8=0.834, M_\nu=0 \}$, and contains additional neutrino cosmologies with total mass $M_\nu=0.1$, $0.2$ and $0.4$ eV equally distributed among the three neutrino species. We use $N_r=100$ realizations for each cosmology, each one with a volume $(1\hgpc)^3$, containing $512^3$ CDM particles and $512^3$ neutrino particles. All simulations have the same $\sigma_8$ value, corresponding to scalar primordial amplitudes $A_s^{M_\nu=0}=2.13 \times 10^{-9}$, $A_s^{M_\nu=0.1 \,\text{eV}}=2.25 \times 10^{-9}$, $A_s^{M_\nu=0.2 \,\text{eV}}=2.40 \times 10^{-9}$ and $A_s^{M_\nu=0.4 \,\text{eV}}=2.74 \times 10^{-9}$. We consider halos with masses in the range $13.1 < \log (M/ h^{-1} M_\odot) < 13.5$ identified with a Friends-of-friends algorithm \cite{Davis:1985rj} over the CDM particles only, with a linking length parameter $b=0.2$. We choose to show results for the redshifts $z=0.5$ and $z=1$, though the case $z=0$ has also been checked, finding similar behaviours.

To compare our analytical model of eq.~\eqref{Pells} against the simulated data we let free nine parameters: four bias parameters $\{b_1, b_2, b_{s^2}, b_{3nl}\}$, three counterterms $\{\alpha_0, \alpha_2,\alpha_4\}$, 
and two stochastic parameters $\{\alpha^{shot}_0-1,\alpha^{shot}_2\}$. Given that we subtracted the Poissonian shot noise $P_{shot}$ from the monopole data, hence the parameter to fit is $\alpha^{shot}_0-1$ instead of $\alpha^{shot}_0$. 
Moreover, for our fittings we only use the diagonal elements of the covariance matrix, because the non-diagonal ones are dominated by statistical noise, even for different multipoles at the same $k$-bin.
In order to sample the parameter space, we run Monte Carlo Markov Chains (MCMC) using the \texttt{emcee} sampler \cite{ForemanMackey:2012ig} and the \texttt{GetDist}  \texttt{Python} package to present the posteriors \cite{Lewis:2019xzd}.
We choose uniform priors over sufficiently wide intervals because apriori all sensible values are equally likely. (Notice that some recent work, which also estimate cosmological parameters, choose Gaussian priors around expected values obtained through $N$-body simulations and co-evolution \cite{DAmico:2019fhj,Nishimichi:2020tvu,Chudaykin:2020aoj}.) 

The first three non-vanishing multipoles ($\ell=0,2,4$) with the best fit parameters for redshifts $z=0.5$ and $z=1$, together with those obtained from the simulations, are depicted in figure \ref{fig:multipoles}. The error bars denote the RMS errors for the $N_r$ realizations on each model. We choose to show the fiducial cosmology $M_\nu=0$ and the massive neutrino $M_\nu= 0.4 \, \text{eV}$, the latter presenting the largest deviation from the massless neutrino cosmology. With the exception of the hexadecapole, which is very noisy, the analytical results lie within the RMS error bars at all wave-number bins. We add that we have found very similar agreement for the rest of massive neutrino cosmology cases: $M_\nu=0.1$ and $0.2$ eV.  In the left panel of figure \ref{fig:b1b2_coev} we plot the posterior distribution contour for the subspace $b_{s^2}$-$b_{3nl}$ at $1\sigma$ and $2\sigma$, together with their values given by co-evolution in EdS, given by eqs.~\eqref{coevbiases} (vertical gray bands, with width given by the dispersion of $b_1$). Although the dispersion in $b_{s^2}$-$b_{3nl}$ is large, the final results are not affected by this: indeed, we also fixed the bias parameters $b_{s^2}$ and $b_{3nl}$ to their co-evolution values before performing the MCMC, and find no differences in the precision of the final result; and importantly, the linear bias estimations remains almost equal. In the right panel of figure \ref{fig:b1b2_coev} we show the contour plots for the subspace $b_{1}$-$b_{2}$. The observed positive correlation is expected from Peak-Background-Split with a Sheth-Tormen (or Press-Schechter) mass function for halos in this mass range \cite{Kaiser:1984sw,Mo:1996cn,Desjacques:2016bnm}. Also, the differences in the large-scale bias is small, consistent with the almost universal halo mass function for cosmologies that have the same amplitude $\sigma_8$ \cite{Villaescusa-Navarro:2013pva}. 
We find the best fit values for the linear local bias $b_1=1.72, 1.71, 1.69, 1.67$, for $M_\nu=0,0.1,0.2,0.4\,\text{eV}$ respectively, when fitting up to $k_{max}=0.25\hmpci$ for redshift $z=0.5$.
Similar values are found for the same halo catalogue in \cite{Aviles:2020cax} using the 2-point correlation function; see Table 1 of that paper. Notice that in that work Lagrangian linear biases are reported, so they are related to the Eulerian linear bias used here as $b_1=1+b_1^L$. 


In figure \ref{fig:ratios} we show the suppression due to the free-streaming for the different neutrino masses and for multipoles $\ell=0$ and $2$. In the left panels we show the power spectrum multipoles multiplied by $k$, and we further multiply by factors $A^{M_\nu=0}/A^{M_\nu}$, such that at large scales the different massive cases differ only because of the different linear biases (which nonetheless are very similar for all the models). Remember that the simulations are normalized to have the same $\sigma_8$, hence by multiplying by the primordial amplitudes we can observe the main differences at scales around and above the free-streaming, which is better to visualize the power spectrum suppression. In the right panels we display the dispersion in the model by sampling the free parameters posterior distribution. \revised{We also show the monopole and quadrupole in the Kaiser power spectrum plus the Poissonian shot noise, $P_K = (b_1 + \mu^2 f)^2 P_L + 1/\bar{n}_X$, in order to observe the scales at which the linear theory departs from the data. Notice here that we are not using any damping factor, so the differences between Kaiser and EFT theories start at quite small $k$.}

We want to discern up to what extent the theory is reliable. To this end, we perform the parameters sampling up to different maximum wave-numbers $k_{max}$ ($=0.25$, $0.35$ and $0.45 \hmpci$). In the left panels of figure \ref{fig:kmax}, we plot the ratios of the analytical predictions to the simulated data for multipoles $\ell=0$ and $2$,  where we use the massive neutrino cosmology with  $M_\nu=0.1\,\text{eV}$ at redsfhift $z=0.5$. We notice that the accuracy of the results are similar for the three cases, all presenting a reasonable fitting to the data up to the different $k_{max}$. However, the estimated values of the free parameters change, which can be very harmful when one estimates the cosmological parameters. Since we are only sampling the nuisances, we put attention to the linear local bias $b_1$, for which we obtain the best fits  $b_1=1.71$, $1.74$ and $1.78$ for $k_{max}= 0.25$, $0.35$ and $0.45$, respectively. These are considerable differences in the estimation of a physical parameter that is intrinsic to the halo catalogue; indeed, we can use different statistics to measure it. In particular, in \cite{Aviles:2020cax} using the correlation function it is found the Lagrangian bias value $b_1^L=0.70$ (corresponding to $b_1=1.70$) for the same catalogue. In the right panel of figure \ref{fig:kmax} we show the halo ZA real space correlation function for the three estimations of $b_1$, together with the data from the simulations; we omit to show the error bars, but they are sufficiently large to encompass the three analytical curves; see figure 7 of \cite{Aviles:2020cax}. We observe that as larger is $k_{max}$, the fitting to the correlation function becomes poorer. We interpret it as a biased estimation of the linear bias; which is particularly important in this context since $b_1$ is very degenerate with the total mass of the neutrinos. 
Two considerations are worthy to mention: First, the correlation function is insensitive to the shot noise because stochastic parameters are uncorrelated to density fields at large scales in configuration space, hence measuring $b_1$ using the correlation function is more direct than with the power spectrum. Second, at large scales the ZA describes correctly the correlation function, including the BAO damping due to the bulk flows of CDM particles. Hence we use linear order in LPT to show the right-panel on figure \ref{fig:kmax}; moreover, moving to higher orders in PT, beyond the ZA, introduces additional Lagrangian bias parameters that are non-locally related to the Eulerian biases, with no general prescription to compare between them.
We believe the main reason for this biased estimation of $b_1$ is the shot noise influence: at moderate non-linear scales the halo power spectrum follows essentially a power law, and it is largely dominated by the $\alpha^{shot}_2 \mu^2 k^2$ term, which becomes tightly constrained by the data for large $k_{max}$. In turn, this constrains even more the constant shot noise, that is degenerate with $b_1$ at large scales. Notice also that at high-$k$ the EFT parameters are quite degenerate with $\alpha^{shot}_0$, because the former scales as $k^2 P_L(k)$ which is close to a constant at small scales; however this degeneracy becomes less important at large scales where the degeneracy between $b_1$ and the constant piece of the shot noise is more relevant. This analysis suggests that the one-loop EFT is not valid for wave-numbers $\gtrsim 0.25 \hmpci$, as expected. Moreover, at the scales where the $(\mu k)^2$ piece of the shot noise dominates, the RSD are driven mainly by stochastic velocities and probably no cosmological information can be extracted \cite{Chen:2020fxs}. In figure \ref{fig:triang} we show triangular posterior contour plots for the shot parameters and $b_1$, for the three analyzed $k_{max}$ showing the degeneracies and tendencies above mentioned. 

Finally, we notice that we may add the next-to-leading order EFT correction introduced in \cite{Ivanov:2019pdj}, with counterterm $\tilde{c}$, that serves for a better modeling of the power spectrum damping along the line-of-sight direction due to the FoG. However, this parameter does not help for a better estimation of $b_1$, while at the same time reaching good accuracy at smaller scales. This is because the introduced functional dependence by $\tilde{c}$ is $k^4 P_L(k)$, which is approximately $\propto k^2$ at high-$k$, and hence degenerate with $\alpha_2^{shot}$.

\revised{In the end, our main interest is to estimate the neutrino mass scale from data. However, as it stands, our theory is quite slow and not suitable for an MCMC cosmological parameter exploration. We have devoted this section to only fit the nuisance parameters, which implies that one would only need to compute the loop integrals once for each model. Indeed, in our preliminary numerical computations, we have also used the EdS kernels to fit the data and find results as good as with our full kernels. However, we have seen in figure \ref{fig:kmax} that a good fit to the data does not imply a correct estimation of parameters. We showed this for the linear bias, but it would not be a complete surprise if a similar situation occurs with the cosmological parameters when using a perturbative theory that is not complete, especially when the data standard deviations are large, and one had to explore, e.g., large neutrino masses, where the EdS kernels and full kernels differ even more. However, to perform a complete MCMC analysis, we need a method that accelerates the computation of loop integrals; such endeavour is the subject of the following section.}

\end{section}

\begin{section}{FFTLog method}\label{sect:FFTlog}

Obtaining loop corrections with the full massive neutrino kernels requires the computation of several integrals of the form $I(k)=\ip K_I(k,p,x)$. Performing these loop calculations is computationally slow, since at each volume element of the integration we need to solve a system of differential equations for second and third order growth functions to find the values of  $\mathcal{A}, \mathcal{B}(k,p,x)$ and their third order counterparts for constructing $K_I(k,p,x)$. The slowness of this procedure precludes the use of efficient parameter sampling algorithms for estimation of cosmological parameters. 

However, the dominant contributions to the loop corrections come from the growth rates instead of these computationally costly functions. The latter calculations mainly play a normalization role, and their effect is similar to how they affect the $\Lambda$CDM kernels, as oppose to the use of EdS kernels.
Therefore, one may use this hierarchy of calculations in order to speed up the one-loop computations. For this purpose, we first define the \texttt{fk} kernels as 
\begin{align}
F_2^\texttt{fk}(\vk_1,\vk_2) &=  F_2(\vk_1,\vk_2)\Big|_{\mathcal{A}=\mathcal{B}=\mathcal{A}^{f_\nu=0}},   \\
G_2^\texttt{fk}(\vk_1,\vk_2) &=  G_2(\vk_1,\vk_2)\Big|_{\mathcal{A}=\mathcal{B}=\mathcal{A}^{f_\nu=0}}, 
\end{align}
and similar for the third order kernels.
In other words, we fix the functions $\mathcal{A}$ and $\mathcal{B}$ to have their large scale ($f_\nu=0$) value $\mathcal{A}^{f_\nu=0}$, which is the same for both functions and coincides with the $\Lambda$CDM value, namely
\begin{equation}
\mathcal{A}^\text{$\Lambda$CDM}(t)=\mathcal{A}^{f_\nu=0}(t)=\mathcal{B}^{f_\nu=0}(t) =\frac{D_\mA(t)}{\frac{3}{7}D^2_+(t)},  
\end{equation}
with
\begin{equation}\label{DAfnu0}
 D_\mA^{f_\nu = 0}(t)= \frac{3}{7}D_+^2(t) +  \frac{4}{7}\left(\frac{d^2 \,}{dt^2} + 2 H \frac{d \,}{dt} - \frac{3}{2} \Omega_m H^2  \right)^{-1} \left[ \frac{3}{2} \Omega_m H^2 
 \left( 1- \frac{f^2}{\Omega_m}\right) \right].
\end{equation}
In the simpler case of EdS, $\Omega_m=1=f$, and the second term in the rhs of the eq.~\eqref{DAfnu0} vanishes, reducing the second order kernels to the well-known EdS ones with $\mA=\mB=1$. For $\Lambda$CDM, or equivalently $f_\nu=0$, and in the case of observationally allowed cosmologies, one obtains numerically $\mA^{f_\nu=0}(z=0) \approx 1.01$.  

In order to asses the accuracy of  simplified kernels, we measure the relative difference of the matter power spectra multipoles 
for $\Lambda$CDM or \texttt{fk} kernels with respect to the computation using the full kernels. That is, 
in figure \ref{fig:KernelsComparison} we plot the variable
\begin{equation}\label{reldiff}
\frac{\Delta P_\ell}{P_\ell} = \frac{P_\ell^\text{$\Lambda$CDM kernels, fk kernels}}{P_{\ell}^\text{Full kernels}}  - 1,  
\end{equation}
for multipoles $\ell=0,2,4$. \revised{In the previous expression the multipoles $P_\ell^\text{fk kernels}$ use the \texttt{fk} kernels, which we remind the reader are obtained by considering the pieces of the full kernels with the scale-dependent growth rates only, while for the pieces that are solution to second order differential equations we assign their massless neutrino values. }
We find that the use of the \texttt{fk} kernels is always very close to the full kernels case (within $0.3 \%$) for $k<0.5\, h\,$Mpc$^{-1}$. In summary, this accuracy of the \texttt{fk} kernels allow us to  use the \textsc{FFTLog} methods \cite{McEwen:2016fjn,Fang:2016wcf,Schmittfull:2016jsw,Schmittfull:2016yqx,Simonovic:2017mhp,Chudaykin:2020aoj} to further accelerate the computation of loop integrals.

 \begin{figure}
 	\begin{center}
 	\includegraphics[width=3.0 in]{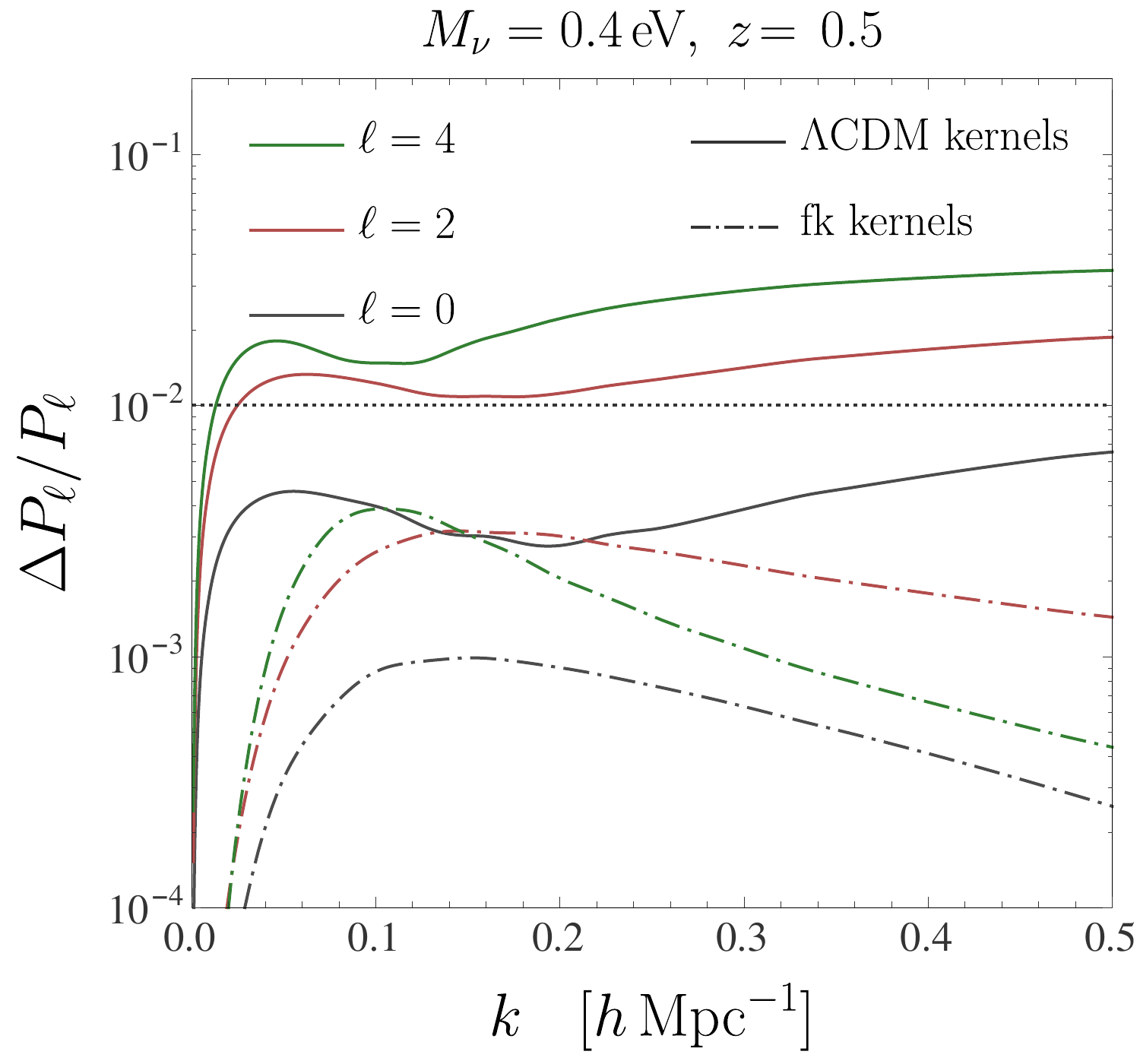}
 	\caption{Relative differences of the multipoles of the redshift-space matter power spectrum when using $\Lambda$CDM kernels (solid lines) and \texttt{fk} kernels (dot-dashed lines) instead of the full kernels, as given by eq.~\eqref{reldiff}. 
 	\label{fig:KernelsComparison}}
 	\end{center}
 \end{figure}

It is not our intention to work out the FFTLog procedure in detail for all the expressions in this work (which will be presented elsewhere), but to show explicitly how this can be done for a particular example. We choose the case of  $P^{22}_{\delta\theta}$, given by the second term in eq.~\eqref{Pdtloop}. In fact, we also show some numerical results for the one-dimensional $P_{\delta\theta}$ and $P_{\theta\theta}$ spectra. For simplicity we assume $\mathcal{A}^{f_\nu=0}=\mathcal{A}^\text{EdS} =1$; though the computation without this restriction is straightforward and it is outlined at the end of this section.

First, we expand the linear power spectrum as a sum of scale invariant spectra with complex powers as \cite{Hamilton:1999uv,McEwen:2016fjn}
\begin{equation} \label{pklfft}
\bar{P}_L(k) = \sum_{m=-N/2}^{N/2} c_m k^{\nu + i \eta_m}, \qquad \eta_m = \frac{N-1}{N}\frac{2 \pi m}{\ln(k_\text{max}/k_\text{min})},    
\end{equation}
where we have split an interval $[k_\text{min},k_\text{max}]$ in $N$ logarithmic spaced  wave-numbers. The coefficients $c_m$ comes from the  discrete log-Fourier transform 
\begin{align} \label{cm}
c_m =  W_m k_\text{min}^{-\nu - i \eta_m} \frac{1}{N}\sum_{l=0}^{N-1} P_L(k_l)\left(\frac{k_l}{k_\text{min}}\right)^{-\nu} e^{- 2 \pi i m l/N},
\end{align}
with the weights $W_m =1$, except for the end points, for which $W_{-N/2}=W_{N/2} =1/2$.
The so-called bias $\nu$ is in principle any real number, but its value is chosen to have a better convergence for loop integrals \cite{Simonovic:2017mhp}.

We need a similar expansion to eq.~\eqref{pklfft} for the linear density-velocity cross-spectra $P^L_{\delta\theta} = (f(k)/f_0) P_L(k)$, namely
\begin{equation} \label{pkldvfft}
\bar{P}^L_{\delta\theta}(k) = \sum_{m=-N/2}^{N/2} c_m^f k^{\nu + i \eta_m},    
\end{equation}
where the coefficients $c_m^f$ are computed with eq.~\eqref{cm} by substituting $P_L(k)$ by $(f(k)/f_0)P_L(k)$.

In the standard method, the kernels are taken to be EdS and these are written in powers of $k$, $p$ and $|\vk-\vp|$. The difficulty in our case is the presence of growth rate functions in $G_2$.  
Hence, we split the integrand of the loop piece of $P^{22}_{\delta\theta}$ into a contribution with a linear growth rate evaluated at the wavenumber $p$ and a further piece evaluated at $|\vk-\vp|$. The splitting results in
\begin{align} \label{P22split}
P^{22}_{\delta\theta} &= 2 \ip F_2^\text{fk}(\vp,\vk-\vp)G_2^\text{fk}(\vp,\vk-\vp) P_L(p) P_L(|\vk-\vp|) \nonumber\\
  &= 2 \ip K^{f_{\vp}}_{\delta\theta}(\vp,\vk-\vp)\frac{f(p)}{f_0} P_L(p) P_L(|\vk-\vp|) \nonumber\\
&\quad + 2 \ip K^{f_{\vk-\vp}}_{\delta\theta}(\vp,\vk-\vp) P_L(p) \frac{f(|\vk- \vp|)}{f_0} P_L(|\vk-\vp|), 
\end{align}
such that the two functions
\begin{align}\label{Kfkmp}
K^{f_{\vk-\vp}}_{\delta\theta}(\vp,\vk-\vp) &=  \frac{k^8}{196 p^4 |\vk-\vp|^4 }-\frac{k^6}{392  p^2 |\vk-\vp|^4}+\frac{3k^6}{196 p^4 |\vk-\vp|^2 }-\frac{9 k^4}{392 |\vk-\vp|^4} \nonumber\\ 
   &\quad +\frac{23 k^4}{784  p^2 |\vk-\vp|^2}-\frac{11 k^4}{784 p^4}+\frac{13 k^2 }{392 p^{-2}|\vk-\vp|^4}-\frac{15 k^2 }{392 p^4|\vk-\vp|^{-2}}  \nonumber\\ 
   &\quad -\frac{5 k^2}{98|\vk-\vp|^2}+\frac{11 k^2}{196 p^2}-\frac{5 }{392 p^{-4}|\vk-\vp|^4}+\frac{25}{784 p^4 |\vk-\vp|^{-4}}   \nonumber\\ 
   &\quad +\frac{5 }{784 p^{-2}|\vk-\vp|^2}-\frac{65}{784 p^2 |\vk-\vp|^{-2}}+\frac{45}{784},     
\end{align}
and
\begin{align}\label{Kfp}
K^{f_{\vp}}_{\delta\theta}(\vp,\vk-\vp) &=  \frac{k^8}{196  p^4|\vk-\vp|^4}+\frac{3 k^6}{196 p^2 |\vk-\vp|^4}-\frac{k^6}{392  p^4|\vk-\vp|^2}-\frac{11 k^4}{784 |\vk-\vp|^4} \nonumber\\
 &\quad +\frac{23k^4}{784 p^2|\vk-\vp|^2 }-\frac{9 k^4}{392 p^4}-\frac{15 k^2 }{392 p^{-2}|\vk-\vp|^4}+\frac{13 k^2 }{392 p^4 |\vk-\vp|^{-2}}  \nonumber\\
 &\quad+\frac{11 k^2}{196|\vk-\vp|^2}-\frac{5 k^2}{98 p^2}+\frac{25 }{784 p^{-4}|\vk-\vp|^4}-\frac{5}{392 p^4 |\vk-\vp|^{-4}} \nonumber\\
 &\quad-\frac{65 }{784 p^{-2}|\vk-\vp|^2}+\frac{5}{784 p^2 |\vk-\vp|^{-2}}+\frac{45}{784},    
\end{align}
are independent of the input cosmology. As a result of the previous expressions, one gets that \begin{equation}
 K^{f_{\vp}}_{\delta\theta}(\vp,\vk-\vp)+K^{f_{\vk-\vp}}_{\delta\theta}(\vp,\vk-\vp)=F_2^\text{EdS}(\vp,\vk-\vp)G_2^\text{EdS}(\vp,\vk-\vp).   
\end{equation}
Since all the summands in the kernels above are of the form $k^{-2(n_1+n_2)}p^{2n_1} |\vk-\vp|^{2 n_2} $ with $n_1,n_2 \in \{-2,-1,0,1,2\}$, it is convenient to define the matrix $f^{f_{\vp}}_{22,\delta\theta}(n_1,n_2)$, with components given by the coefficient of each term in eq.~\eqref{Kfp}. Similarly, one constructs the matrix $f^{f_{\vk-\vp}}_{22,\delta\theta}(n_1,n_2)$ using eq.~\eqref{Kfkmp}. 
Therefore, using these matrices and with the aid of the discrete Fourier transforms [eqs.~\eqref{pklfft} and \eqref{pkldvfft}], the $P_{22}$ of eq.~\eqref{P22split} can be approximated by
\begin{align}
P^{22}_{\delta\theta}(k) &= 2 \sum_{m_1,m_2} c^{f}_{m_1} c_{m_2} \sum_{n_1,n_2=-2}^2 f^{f_{\vp}}_{22,\delta\theta}(n_1,n_2) k^{-2(n_1+n_2)} \ip \frac{1}{p^{2\nu_1-2 n_1} |\vk-\vp|^{2\nu_2-2 n_2}} \nonumber\\
& +  2 \sum_{m_1,m_2} c_{m_1} c^{f}_{m_2} \sum_{n_1,n_2=-2}^2 f^{f_{\vk-\vp}}_{22,\delta\theta}(n_1,n_2) k^{-2(n_1+n_2)} \ip \frac{1}{p^{2\nu_1-2 n_1} |\vk-\vp|^{2\nu_2-2 n_2}}, 
\end{align}
with
\begin{equation}
\nu_1 = -\frac{1}{2}( \nu + i \eta_{m_1}) , \qquad \nu_2 = -\frac{1}{2}( \nu + i \eta_{m_2}).   
\end{equation}
To further simplify the previous expression, we define
\begin{align}
M_{22,\delta\theta}^{f_{\vp}}(\nu_1,\nu_2) &=  2\sum_{n_1,n_2=-2}^2  f^{f_{\vp}}_{22,\delta\theta}(n_1,n_2) I (\nu_1-n_1,\nu_2-n_2), \label{M22fp} \\
M_{22,\delta\theta}^{f_{\vk-\vp}}(\nu_1,\nu_2) &=  2\sum_{n_1,n_2=-2}^2 f^{f_{\vk-\vp}}_{22,\delta\theta}(n_1,n_2) I (\nu_1-n_1,\nu_2-n_2), \label{M22fkmp}
\end{align}
and  \cite{Smirnov:1991jn,Scoccimarro:1996jy}
\begin{align}
I(a,b) \equiv k^{-3 + 2(a+ b)}\ip \frac{1}{p^{2a} |\vk-\vp|^{2b}} 
= \frac{1}{8 \pi^{3/2}} \frac{\Gamma(\frac{3}{2} - a)\Gamma(\frac{3}{2} - b)\Gamma(a+b-\frac{3}{2})}{\Gamma(a)\Gamma(b)\Gamma(3-a-b)}.   
\end{align}
Using the above equations, one can write the approximation for $P^{22}_{\delta\theta}$ as
\begin{align}
\bar{P}^{22}_{\delta\theta}(k) &= k^3 \sum_{m_1,m_2} c_{m_1}^f  k^{-2 \nu_1} \, M_{22,\delta\theta}^{f_{\vp}}(\nu_1,\nu_2) \,  c_{m_2} k^{-2 \nu_2} \nonumber\\ &\quad 
+ k^3 \sum_{m_1,m_2} c_{m_1}  k^{-2 \nu_1} \, M_{22,\delta\theta}^{f_{\vk-\vp}}(\nu_1,\nu_2) \,  c^f_{m_2} k^{-2 \nu_2}.
\end{align}

 \begin{figure}
 	\begin{center}
 	\includegraphics[width=6.0 in]{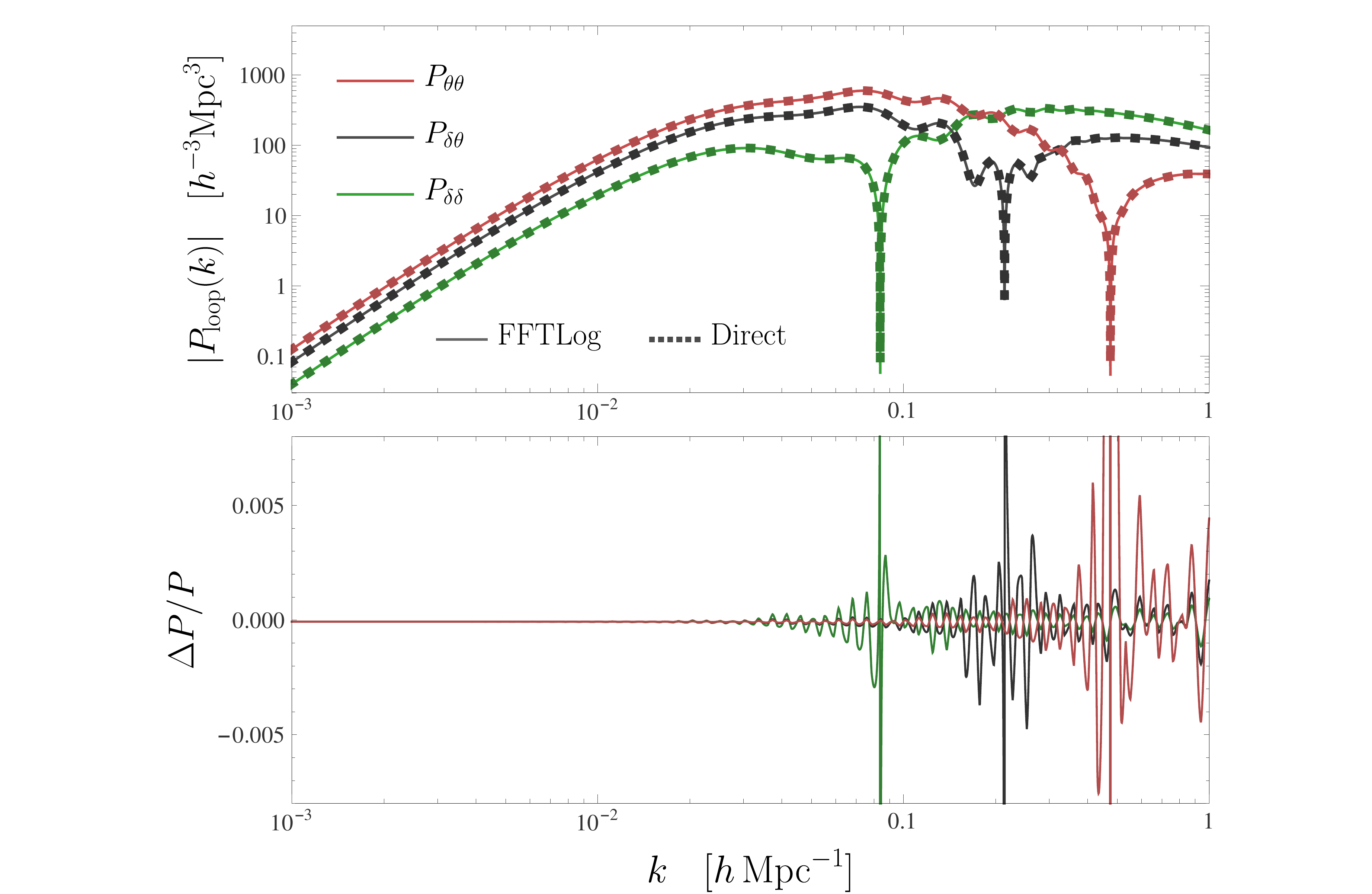}
 	\caption{Comparison of the \textsc{FFTLog} and Direct computations for the one-loop contributions to $P_{\delta\delta}$, $P_{\delta\theta}$ and $P_{\theta\theta}$. The parameters are $k_\text{min}=10^{-6} \,\hmpci$, $k_\text{max}=10  \,\hmpci$, $N=256$ and bias $\nu=-0.1$. 
 	\label{fig:fftlog}}
 	\end{center}
 \end{figure}

We notice that the $M_{22}$ matrices defined in eqs.~\eqref{M22fp} and \eqref{M22fkmp} are not symmetric. However, the interchange $\vp \rightarrow \vk-\vp$ symmetry, which explicitly holds in the first line of eq.~\eqref{P22split}, implies
$M_{22,\delta\theta}^{f_{\vp}}(\nu_1,\nu_2)=M_{22,\delta\theta}^{f_{\vk-\vp}}(\nu_2,\nu_1)$. This equality permits us to reduce the previous expression to 
\begin{align}
\bar{P}^{22}_{\delta\theta}(k) &= 2 k^3 \sum_{m_1,m_2} c_{m_1}^f  k^{-2 \nu_1} \, M_{22,\delta\theta}^{f_{\vp}}(\nu_1,\nu_2) \,  c_{m_2} k^{-2 \nu_2}.
\end{align}
Note that the EdS matrix is recovered as $M_{22,\delta\theta}^\text{EdS}= M_{22,\delta\theta}^{f_{\vp}}+M_{22,\delta\theta}^{f_{\vk-\vp}}$.

The computation of the rest of loop contributions to construct the one-dimensional spectra of eqs.~\eqref{Pddloop}, \eqref{Pdtloop} and \eqref{Pttloop} is similar, but lengthy so we do not show them here. 
In figure \ref{fig:fftlog} we show the results for $P_{\delta\delta}$, $P_{\delta\theta}$ ($\ni P^{22}_{\delta\theta}$) and $P_{\theta\theta}$ computed using \textsc{FFTLog} and the \textsc{Direct} integration methods. To enhance the effects of massive neutrinos we work the case of $M_\nu=0.4\,\text{eV}$, for which the introduction of the scale dependent growth rate is more important. In conclusion, we show that the FFTLog method, using a bias of $\nu=-0.1$,\footnote{After a few tests we found this number yield better convergence. For bias $\nu>-1$ one has to add the UV limits of the $P_{13}$ functions.}
works well to a very high accuracy, smaller than the $0.5\%$ over the interval $\,0.001 <k< 1 \hmpci$. 

Finally, let us discuss how to introduce the correct time dependence of the $\mA$ function that we have so far neglected. Notice that the coefficients in the kernels of eqs.~\eqref{Kfkmp} and \eqref{Kfp} would become multiplied by factors $\epsilon^m d^n$ (with $m+n=0,1,2$), where 
\begin{equation}
    \epsilon = \mA^{f_\nu=0}(t)-1 \lesssim 0.01, \qquad \text{and} \qquad d=\frac{\dot{\mA}^{f_\nu=0}(t)}{ f_0H}  \lesssim 0.01.
\end{equation}
As a result, the $M_{22}$ matrices can be split as
\begin{equation}
M_{22} = M_{22}\big|_{\epsilon=d=0} + \epsilon \Big[M_{22}-M_{22}\big|_{\epsilon=d=0}\Big]_{\epsilon=1,d=0}  + d \Big[M_{22}-M_{22}\big|_{\epsilon=d=0}\Big]_{\epsilon=0,d=1}, 
\end{equation}
where we neglect terms of order $\epsilon^m d^n$ with $m+n=2$. 

This method can also be applied to find the $\Lambda$CDM model with the exact time dependence (or general models with non-clustering dark energy), at the cost of introducing two additional matrices. In such a case, one would have $M_{22}^\text{$\Lambda$CDM}\big|_{\epsilon=d=0} =  M^\text{EdS}_{22}$.

\end{section}

\begin{section}{Summary and Conclusions}\label{sect:conclusions}

Despite neutrinos being non-relativistic particles at late times and behaving analogous to other matter components, rather than radiation, they are not cold components due to their large velocity dispersions. Depending on the scales of interest, neutrinos present two distinct behaviors. In regions below the free-streaming scale, a Jeans-like mechanism with a pressure supported by the neutrino thermal velocity, prevents their confinement. In contrast, for scales well above this scale, 
they cluster in the same way as the CDM component. 
As a consequence, matter statistics become suppressed below the free-streaming when compared to the equivalent massless neutrino scenario.

In the present work, we built upon the work of refs.~\cite{Aviles:2020cax} and \cite{Aviles:2020wme} to construct an Eulerian Perturbation Theory (EPT) for the Large-Scale clustering in massive neutrino cosmologies.
The most important assumption we made is given by eq.~\eqref{alphaApprox}, where the non-linear neutrino density is approximated by the non-linear $cb$ density fluctuation times the ratio of the neutrinos to $cb$ transfer functions. As a result, one obtains a theory that exhibits a scale-dependent gravitational strength, which has been partially motivated by the recent developments in PT for modified gravity models (e.g. \cite{Koyama:2009me,Aviles:2017aor}). Of particular relevance for this work are the findings of \cite{Aviles:2020wme}, where a generalization of the kernels beyond the EdS approximation were developed for the density-weighted velocity moments 
(see \cite{Vlah:2018ygt}). 

We do not try to obtain the EPT kernels directly from the continuity, Euler and Poisson equations, but instead, we map the Lagrangian Perturbation Theory (LPT) kernels of \cite{Aviles:2020cax} into the Eulerian frame. This approach is considerably more straightforward, and for theories with additional scales, the computation of LPT kernels is generally simpler than for the EPT counterparts. However, and for the sake of completeness, we show how to derive the second order kernels directly in appendix \ref{app:2EPTKernels}, matching the other approach. Moreover, our EPT kernels are well behaved in the UV and IR limits, and reduce to the $\Lambda$CDM expressions when using the large scale limit in the function $A(k,t)$, introduced in eq.~\eqref{Akt}.

In order to asses the accuracy of our predictions we contrast our modelling with synthetic data. In particular, we compare the first three non-vanishing multipoles of the power spectrum to a halo catalogue from the \textsc{Quijote} simulations. We find good agreement up to the commonly expected validity scale in the EFT ($k\sim 0.25 \hmpci$). We can extend to higher wave-numbers with increasing precision, but by doing so we are incorrectly estimating the linear bias. One need to remember that this bias is physical, intrinsic to the halo catalogue and degenerate with the neutrino masses, hence this is potentially harmful for the sum of neutrino mass estimations.

Given the small discrepancies of the power spectrum predictions between EdS and full kernels for realistic neutrino masses, the functional liberty allowed by counterterms, biasing and stochasticity, as well as the expected precision of current and upcoming galaxy surveys, one may be skeptical about the necessity of using a more complicated theory, which is also much more demanding computationally. However, recent analyses of 
large scale structure BOSS data \cite{Ivanov:2019pdj,Palanque-Delabrouille:2019iyz} have shown that the constraining power on the neutrino masses of surveys without adding further data-sets is limited, with the need of sampling parameters up to large masses (of a few eV), where the use of EdS kernels is questionable. Another reason to use neutrino kernels is that higher order statistics may break degeneracies between the galaxy bias and the neutrino masses. In particular, recent results for the redshift space bispectrum show that the use of a proper description of $G_2$, away from the EdS approximation, may help breaking degeneracies between $b_1$ and $f_\nu$ \cite{Kamalinejad:2020izi}. In spite of these motivations, the computation of loop corrections with full kernels is quite expensive, which makes the use of this theory unviable for routinely cosmological parameter estimation. For this reason, we reduce the information content of the kernels, but keep its main feature, the inherited ---from linear theory--- effect of having a scale-dependent growth rate. 
The advantage of these reduced kernels is twofold: first, there is no need of solving differential equations at each step of the loop correction quadratures; and second, these kernels are suitable for an FFTLog method. We make some developments in the latter and find very good agreement with the direct integration method for the three first density-weighted velocity moments. 

\revised{On the other hand, we have discussed at the end of section \ref{sect:results} how the PT--EFT theory stops being that accurate at around $k = 0.25 \,h\,\text{Mpc}^{-1}$. Actually, one is able to fit the data with high precision at higher wavelengths, but it results in an inaccurate estimation of the linear bias. This is better illustrated in figure \ref{fig:kmax} where both, the correlation function and the power spectrum are fit to the data. It would not be a surprise if this same biased estimation when using large wavelengths happens with the cosmological parameters. 
In our early numerical results we found that by fitting the nuisance parameters with the use of EdS kernels, instead of the exact ones, one obtains good results, with slightly different best-fitting nuisance parameters. But in the same line of thought as above, we can speculate further and conceive that the use of EdS kernels in fitting the cosmological parameters may lead to the wrong estimation of parameters, particularly when the standard deviations of measurements are large and one has to explore large neutrino masses. Therefore, exploring the whole space of parameters with the use of MCMC algorithms may turn out to be very useful. In this context, the full FFTLog analysis is a natural avenue to carry on this exploration and we leave it for future work.}

\end{section}

\acknowledgments

We thank Hern\'an E.~Noriega and Mario A.~Rodriguez-Meza for useful discussions, together with the Instituto Avanzado de Cosmología A.~C. and the DCI-UG DataLab for academic and computational resources. AA and GN acknowledge the support of CONACyT, specially through the projects 283151, 286897 and Ciencia de Frontera No.~102958. GN also appreciates the grants of UG-DAIP. AB was supported by the Fermi Research Alliance, LLC under Contract No. DE-AC02-07CH11359 with
the U.S. Department of Energy, and the U.S. Department of Energy (DOE) Office of Science Distinguished Scientist Fellow Program

\appendix

\begin{section}{Direct computation of second order Eulerian kernels}\label{app:2EPTKernels}


In this appendix we find the second order EPT kernels directly from the fluid equations and show that these coincide with those obtained with the mapping between Lagrangian and Eulerian frames used in section \ref{sect:LPTtoSPT}. 
The continuity and Euler equations subjected to the scale dependent gravitational strength $A(k)$ are
\begin{align}
\frac{1}{H}  \frac{\partial\delta(\vk)}{\partial t} - f_0 \theta(\vk) &=  f_0  \ikk \alpha(\vk_1,\vk_2) \theta(\vk_1) \delta(\vk_2), \label{FScontEq}\\
 \frac{1}{H}  \frac{\partial f_0 \theta(\vk)}{\partial t} + \left( 2+ \frac{\dot{H}}{H^2}\right) f_0 \theta(\vk)
&- \frac{A(k)}{H^{2}} \delta(\vk)   =   f_0^2 \ikk \beta(\vk_1,\vk_2) \theta(\vk_1) \theta(\vk_2), \label{FSEulerEq}
\end{align}
with
\begin{equation}
 \alpha(\vk_1,\vk_2) = 1+\frac{\vk_{1}\cdot\vk_2}{k_1^2},  \qquad  \beta(\vk_1,\vk_2) = \frac{k_{12}^2(\vk_1\cdot\vk_2)}{2 k_1^2 k_2^2}, 
\end{equation}
where we used the Poisson equation $(k/a)^2\Phi = -A(k,t) \delta$ and the definition of the rescaled velocity divergence $\theta = -i k_i v^i / (aHf_0)$ in terms of the peculiar velocity $v^i$. 

We introduce the EPT kernels through
\begin{align}
 \delta^{(n)} (\vk,t) &= \underset{\vk_{1\cdots n}= \vk}{\int} F_n(\vk_1,\cdots,\vk_n;t) \delta_L(\vk_1,t) \cdots \delta_L(\vk_n,t), \nonumber\\
 \theta^{(n)} (\vk,t) &= \underset{\vk_{1\cdots n}= \vk}{\int} G_n(\vk_1,\cdots,\vk_n;t) \delta_L(\vk_1,t) \cdots \delta_L(\vk_n,t).
\end{align}
Hence, at first order, 
\begin{align}
 F_1(\vk) &=1, \qquad \text{and} \qquad
 G_1(\vk) = \frac{f(k)}{f_0},
\end{align}
with the growth rate $f(k,t) =d\ln D_+ / d \ln a$, $f_0=f(k\rightarrow 0)$, and the linear growth function solving the differential equation $\ddot{D}_+ +  2H \dot{D}_+ = A(k) D_+$, with appropriate initial conditions that pick out the fastest growing solution.

To second order, the fluid equations are
\begin{align}
 H^{-1}\frac{\partial\delta^{(2)}(\vk)}{\partial t} - f_0\theta^{(2)}(\vk) &=  f_0 \ikk \alpha(\vk_1,\vk_2) \theta^{(1)}(\vk_1) \delta^{(1)}(\vk_2) \nonumber \\
 &=\frac{1}{2}\ikk \big[\alpha(\vk_1,\vk_2)f(k_1) + \alpha(\vk_2,\vk_1)f(k_2) \big]\delta^{(1)}(\vk_1) \delta^{(1)}(\vk_2) \label{ContFS2},
\end{align}
\begin{align} 
 H^{-1}\frac{\partial f_0 \theta^{(2)}(\vk)}{\partial t} + f_0 \left( 2 + \frac{\dot{H}}{H^2}\right) \theta^{(2)}(\vk)
& - \frac{A(k)}{H^{2}} \delta^{(2)}(\vk)  =   f_0^2 \ikk \beta(\vk_1,\vk_2) \theta^{(1)}(\vk_1) \theta^{(1)}(\vk_2), \nonumber\\ 
&=    \ikk \beta(\vk_1,\vk_2) f(k_1)f(k_2) \delta^{(1)}(\vk_1) \delta^{(1)}(\vk_2) \label{EulerFS2},
\end{align}
where we have used $\theta^{(1)}(\vk) = (f(k)/f_0)\delta^{(1)}(\vk)$, and inside the integral of the rhs of eq.~\eqref{ContFS2} we have symmetrized over. 

The second order density fluctuation and velocity fields are
\begin{align}
 \delta^{(2)} (\vk) &= \underset{\vk_{12}= \vk}{\int} F_2(\vk_1,\vk_2) D_+(\vk_1,t)D_+(\vk_2,t)\delta_0(\vk_1) \delta_0(\vk_2), \nonumber\\
 \theta^{(2)} (\vk) &= \underset{\vk_{12}= \vk}{\int} G_2(\vk_1,\vk_2) D_+(\vk_1,t)D_+(\vk_2,t)\delta_0(\vk_1) \delta_0(\vk_2)
\end{align}
with $\delta_0(\vk) = \delta^{(1)}(\vk,t_0)$. Inserting these expressions into eqs.~\eqref{ContFS2} and \eqref{EulerFS2},
\begin{align}
 &\frac{1 }{H D_1D_2} \frac{d \,\,}{d t}(F_2 D_1D_2) - f_0 G_2 = 
            \frac{1}{2} (\alpha_{12}f_1+\alpha_{21}f_2) \\
 &\frac{1 }{H D_1D_2} \frac{d \,\,}{d t}(f_0 G_2 D_1D_2)     + \left( 2 + \frac{\dot{H}}{H^2}\right) f_0 G_2
   -\frac{A(k)}{H^2}F_2   = \beta_{12} f_1 f_2,
\end{align}
with $f_{1,2}=f(k_{1,2})$, $D_{1,2}=D_+(\vk_{1,2},t)$, $\alpha_{12}=\alpha(\vk_1,\vk_2)$, $\alpha_{21}=\alpha(\vk_2,\vk_1)$ and $\beta_{12}=\beta(\vk_1,\vk_2)$. 
We rewrite the above equations as
\begin{align}
&\frac{1}{H}  \frac{d F_2}{d t} + F_2(f_1+f_2) - f_0 G_2 =  \frac{1}{2} (\alpha_{12}f_1+\alpha_{21}f_2), \label{Ceq3}\\
&\frac{1}{H}  \frac{d f_0 G_2}{d t} + f_0 G_2(f_1+f_2) + \left( 2 + \frac{\dot{H}}{H^2}\right) f_0 G_2
   -\frac{A(k)}{H^2}F_2   = \beta_{12} f_1 f_2. \label{Eeq3}
\end{align}

Taking the time derivative of eq.~\eqref{Ceq3} and using eq.~\eqref{Eeq3} we obtain a second order equation for $F_2$,
\begin{align}
& \frac{1}{H^2}\ddot{F}_2 + \frac{2}{H}(1+f_1+f_2)\dot{F}_2 + \left[ \frac{1}{H}(\dot{f}_1 + \dot{f}_2) 
 + (f_1+f_2)\left(f_1+f_2 + 2 + \frac{\dot{H}}{H^2} \right) - \frac{A(k)}{H^2}\right] F_2 = \nonumber\\
& \qquad \frac{1}{2 H} (\alpha_{12}\dot{f}_1 + \alpha_{21}\dot{f}_2) +  \frac{1}{2 } (\alpha_{12}f_1 + \alpha_{21}f_2)\left(f_1+f_2 + 2 + \frac{\dot{H}}{H^2} \right)
+ \beta_{12}f_1f_2,  \label{ddotF}
\end{align}
Now, the growth rate $f(k_1)$ evolves as
\begin{equation}\label{dotfeq}
  \dot{f}_1 = \frac{A(k_1)}{H} - H \left(2+\frac{\dot{H}}{H^2} \right)f_1  - H f_1^2,
\end{equation}
and an equivalent expression for $f_2$. Substituting it in eq.~\eqref{ddotF},
\begin{align}\label{ddotF2}
& \frac{1}{H^2} \ddot{F}_2 + \frac{2}{H}(1+f_1+f_2)\dot{F}_2 + \left[ 2f_1 f_2 + \frac{A(k_1)+A(k_2) - A(k)}{H^2} \right] F_2
 =  \nonumber\\ &\qquad \frac{1}{2}\alpha_{12}\frac{A(k_1)}{H^2} + \frac{1}{2}\alpha_{21}\frac{A(k_2)}{H^2}
  + \frac{1}{2}f_1f_2(\alpha_{12}+\alpha_{21}) + \beta_{12}f_1f_2.
\end{align}
Now, let us define a second order growth function as
\begin{equation}
 D^{(2)}(\vk_1,\vk_2,t) \equiv D_{12} \equiv 2 D_1 D_2 F_2 - \chi_{12},
\end{equation}
and
\begin{align}
\chi_{12} &\equiv \alpha_{12}+\alpha_{21} - \gamma_{12}, \qquad \text{with} \qquad
\gamma_{12} \equiv 1 - \frac{(\vk_1\cdot\vk_2)^2}{k_1^2 k_2^2}. 
\end{align}
We will now find a differential equation for $D_{12}$.
First, the second order $F_2$ kernel is
\begin{equation}\label{F2ansatz}
 F_2 = \frac{D_{12}}{2D_1D_2} + \frac{1}{2}\chi_{12}.
\end{equation}

Now, taking time derivatives of the above equation, 
\begin{equation}
\frac{1}{H}\dot{F}_2 = \frac{1}{2D_1D_2}\left( \frac{1}{H}\dot{D}_{12} - D_{12}(f_1+f_2) \right),
\end{equation}

\begin{align}
  \frac{1}{H^2}\ddot{F}_2 &= \frac{1}{2D_1D_2}\Bigg[ \frac{1}{H^2}\ddot{D}_{12} - \frac{2}{H}(f_1+f_2)\dot{D}_{12} \nonumber\\
  &\quad \qquad +D_{12}\left( 2 (f_1^2 + f_2^2 + f_1f_2 +f_1+f_2)  - \frac{1}{H^2}(A(k_1)+ A(k_2))\right)\Bigg],
\end{align}
where we used eq.~\eqref{dotfeq} and $\ddot{D} + 2H \dot{D} = A(k)D$.
Substituting the above equations into eq.~\eqref{ddotF2},
\begin{align}
 \ddot{D}_{12} + 2H\dot{D}_{12} - A(k)D_{12} &= \Bigg[ A(k)  
 + (A(k)-A(k_2)) \frac{\vk_1 \cdot \vk_2}{k^2_1}
 + (A(k)-A(k_1)) \frac{\vk_1 \cdot \vk_2}{k^2_2} \nonumber\\ 
 &\quad \qquad - (A(k_1)+A(k_2) - A(k))\frac{(\vk_1\cdot\vk_2)^2}{k_1^2k_2^2}  \Bigg] D_1 D_2.
\end{align}
Hence, using eq.~\eqref{F2ansatz} we obtain
\begin{align}
F_2(\vk_1,\vk_2) &= \frac{1}{2} + \frac{3}{14}\A + \left( \frac{1}{2} - \frac{3}{14}\B  \right)   \frac{(\vk_1\cdot\vk_2)^2}{k_1^2 k_2^2}
        + \frac{\vk_1\cdot\vk_2}{2 k_1k_2} \left(\frac{k_2}{k_1} + \frac{k_1}{k_2} \right), \label{F2_kernelapp}
\end{align}
and from eq.~\eqref{Ceq3},
\begin{align}
G_2(\vk_1,\vk_2) &= \frac{3\A(f_1+f_2) + 3 \dot{\A}/H }{14 f_0} +
\left(\frac{f_1+f_2}{2 f_0} - \frac{3\B(f_1+f_2) + 3 \dot{\B}/H }{14 f_0}\right) \frac{(\vk_1\cdot\vk_2)^2}{k_1^2 k_2^2} \nonumber\\
&\quad + \frac{\vk_1\cdot\vk_2}{2 k_1k_2} \left( \frac{f_2}{f_0}\frac{k_2}{k_1} + \frac{f_1}{f_0}\frac{k_1}{k_2} \right), \label{G2_kernelapp}
\end{align}
with $\mA$ and $\mB$ given by
\begin{equation} \label{AandBdef}
 \mA(\vk_1,\vk_2,t) = \frac{7 D^{(2)}_{\mA}(\vk_1,\vk_2,t)}{3 D_{+}(k_1,t)D_{+}(k_2,t)}, 
 \qquad \mB(\vk_1,\vk_2,t) = \frac{7 D^{(2)}_{\mB}(\vk_1,\vk_2,t)}{3 D_{+}(k_1,t)D_{+}(k_2,t)},
\end{equation}
with second order growth functions $D^{(2)}_{\mA,\mB}$ the solutions to the Green problem:
\begin{align}
D^{(2)}_{\mA} = \big(\T - A(k)\big)^{-1}\Big[A(k) &+ (A(k)-A(k_1) ) \frac{\vk_1\cdot \vk_2}{k_2^2} \nonumber\\
                                                  &+(A(k)-A(k_2) ) \frac{\vk_1\cdot \vk_2}{k_1^2} \Big]  D_{+}(k_1)D_{+}(k_2), \label{DAeveq} \\
D^{(2)}_{\mB} = \big(\T - A(k)\big)^{-1} \Big[A(k_1) &+ A(k_2) - A(k) \Big]  D_{+}(k_1)D_{+}(k_2), \label{DBeveq}
\end{align}
with $k=|\vk_1 + \vk_2|$, and \cite{Matsubara:2015ipa}
\begin{equation}
\T = \frac{d^2 \,}{dt^2} + 2 H \frac{d \,}{dt}.    
\end{equation}
From eqs.~\eqref{F2_kernelapp} and \eqref{G2_kernelapp} one recovers eqs.~\eqref{F2_kernelleg} and \eqref{G2_kernelleg}, obtained from LPT.
And by inverting eqs.~\eqref{LPTtoF2} and \eqref{LPTtoG2}, one obtains the LPT kernels  
\begin{align}
\Gamma_2(\vk_1,\vk_2) &= \A -\B \frac{(\vk_1\cdot\vk_2)^2}{k_1^2 k_2^2}, \\ 
\Gamma_2^f (\vk_1,\vk_2) &= \left(\A -\B \frac{(\vk_1\cdot\vk_2)^2}{k_1^2 k_2^2} \right)\frac{f(k_1) + f(k_2)}{2 f_0} + \frac{1}{2 H f_0} \left(\dot{\A} -\dot{\B} \frac{(\vk_1\cdot\vk_2)^2}{k_1^2 k_2^2} \right),
\end{align}
which are those found in \cite{Aviles:2020cax} by a direct LPT approach.

\end{section}

 \bibliographystyle{JHEP}  
 \bibliography{bib.bib}

\end{document}